\documentclass[sigconf,screen]{acmart}

\usepackage{dblfloatfix}
\usepackage{multirow}
\usepackage{booktabs}
\usepackage{array}
\usepackage{algorithm}
\usepackage{eucal}
\usepackage{wrapfig}
\usepackage{datetime}
\usepackage{cleveref}
\crefname{section}{§\hspace{-2pt}}{§§}
\Crefname{section}{§}{§§}
\usepackage{enumitem}
\usepackage[noend]{algpseudocode}
\newcolumntype{L}[1]{>{\raggedright\let\newline\\\arraybackslash\hspace{0pt}}m{#1}}
\newcolumntype{C}[1]{>{\centering\let\newline\\\arraybackslash\hspace{0pt}}m{#1}}
\newcolumntype{R}[1]{>{\raggedleft\let\newline\\\arraybackslash\hspace{0pt}}m{#1}}
\usepackage{tikz}
\newcommand*\circled[1]{\tikz[baseline=(char.base)]{
            \node[shape=circle,fill,inner sep=1pt] (char) {\textcolor{white}{#1}};}}
\makeatletter
\def\thickhline{%
  \noalign{\ifnum0=`}\fi\hrule \@height \thickarrayrulewidth \futurelet
   \reserved@a\@xthickhline}
\def\@xthickhline{\ifx\reserved@a\thickhline
               \vskip\doublerulesep
               \vskip-\thickarrayrulewidth
             \fi
      \ifnum0=`{\fi}}
\makeatother
\newlength{\thickarrayrulewidth}
\usepackage{pifont}

\newcommand{\xmark}{\ding{54}}
\DeclareMathOperator*{\argmax}{argmax} 
\usepackage{hyperref}
\hypersetup{
    colorlinks=true,
    linkcolor=blue,
    filecolor=magenta,      
    urlcolor=cyan,
    pdftitle={Overleaf Example},
    pdfpagemode=FullScreen,
}

\newcommand{\shellcmd}[1]{\\\indent\indent\texttt{\footnotesize\$ #1}}

\newcommand{\rbc}[1]{{#1}}
\newcommand{\rbcb}[1]{{#1}}
\newcommand{\rbcc}[1]{{#1}}
\newcommand{\rbcd}[1]{{#1}}

\AtBeginDocument{%
  \providecommand\BibTeX{{%
    \normalfont B\kern-0.5em{\scshape i\kern-0.25em b}\kern-0.8em\TeX}}}

\setcopyright{acmcopyright}
\copyrightyear{2021}
\acmYear{2021}
\setcopyright{acmlicensed}\acmConference[MICRO '21]{MICRO-54: 54th Annual IEEE/ACM International Symposium on Microarchitecture}{October 18--22, 2021}{Virtual Event, Greece}
\acmBooktitle{MICRO-54: 54th Annual IEEE/ACM International Symposium on Microarchitecture (MICRO '21), October 18--22, 2021, Virtual Event, Greece}
\acmPrice{15.00}
\acmDOI{10.1145/3466752.3480114}
\acmISBN{978-1-4503-8557-2/21/10}

\begin{document}

\fancypagestyle{firstpage}
{
    \fancyhead{}
    \footskip=30pt
	\fancyfoot{}
    \fancyfoot[C]{\thepage}
}
\fancypagestyle{firstpage}
{
    \fancyhead{}
    \begin{tikzpicture}[remember picture,overlay]
    \node [xshift=142mm,yshift=-14mm]
    at (current page.north west) {\href{https://www.acm.org/publications/policies/artifact-review-and-badging-current}{\includegraphics[width=2.2cm]{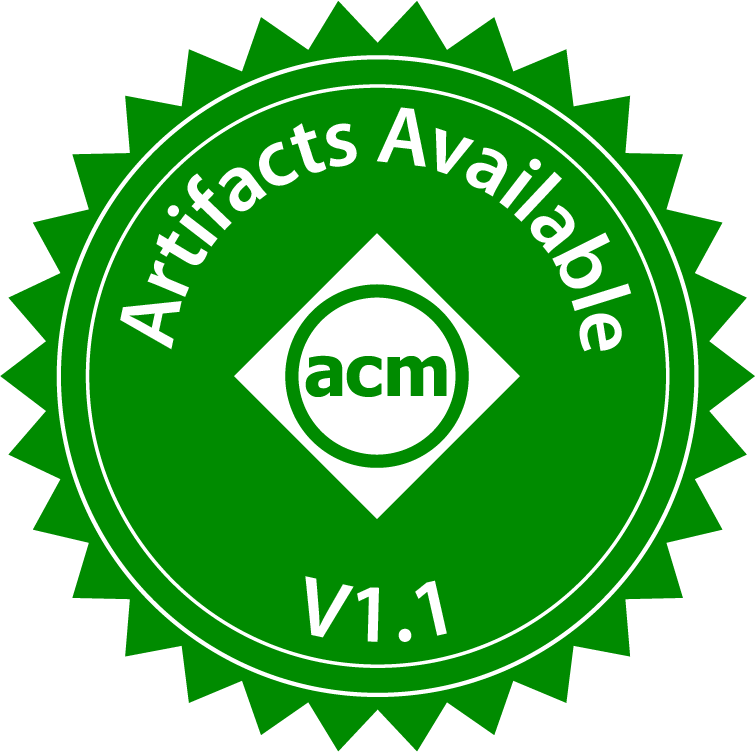}}} ;
    \node [xshift=165mm,yshift=-14mm]
    at (current page.north west) {\href{https://www.acm.org/publications/policies/artifact-review-and-badging-current}{\includegraphics[width=2.2cm]{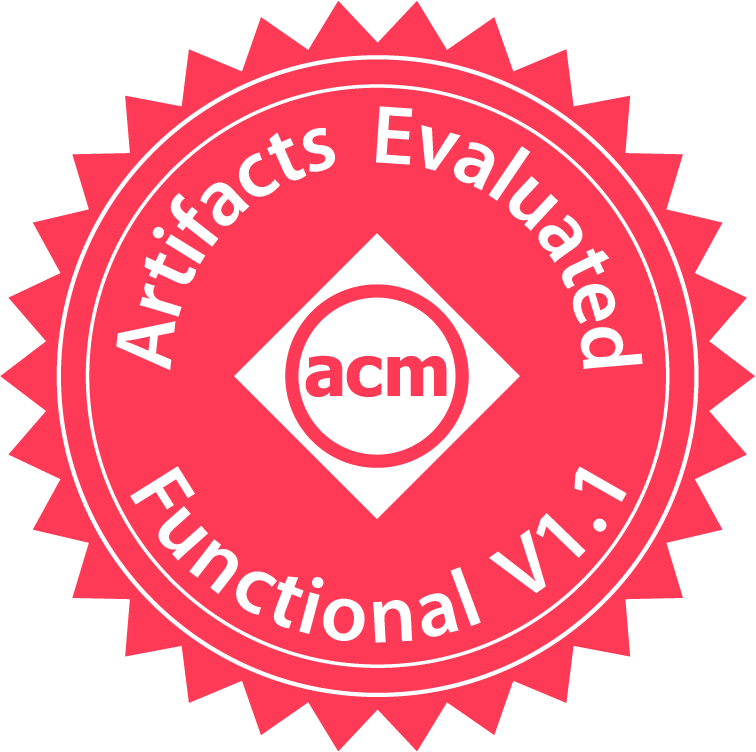}}} ;
    \node [xshift=188mm,yshift=-14mm]
    at (current page.north west) {\href{https://www.acm.org/publications/policies/artifact-review-and-badging-current}{\includegraphics[width=2.2cm]{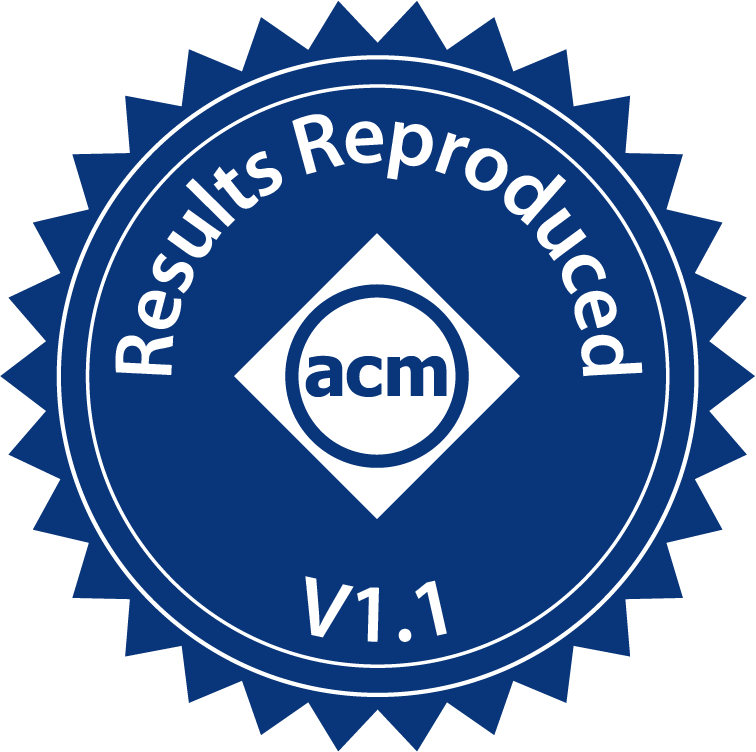}}} ;
\end{tikzpicture}

  \renewcommand{\headrulewidth}{0pt}
  \pagenumbering{arabic}
  \fancyfoot[C]{\large\thepage}
}

\title{Pythia: A Customizable Hardware Prefetching Framework \\Using Online Reinforcement Learning}


\author{Rahul Bera$^1$ \hspace{1em} Konstantinos Kanellopoulos$^1$ \hspace{1em} Anant V. Nori$^2$ \hspace{1em} Taha Shahroodi$^{3,1}$ \\\vspace{0.2em} \hspace{-1em} Sreenivas Subramoney$^2$ \hspace{1em} Onur Mutlu$^1$}
\affiliation{
    \vspace{0.5em}
    \institution{$^1$ETH Zürich \hspace{1em} $^2$Processor Architecture Research Labs, Intel Labs \hspace{1em} $^3$TU Delft}
    \country{}
}

\renewcommand{\shortauthors}{R. Bera and K. Kanellopoulos, et al.}

\begin{abstract}
\rbc{Past research} has proposed numerous hardware prefetching techniques, most of which rely on exploiting one specific type of program context information (e.g., program counter, \rbc{cacheline address}, \rbc{or} delta between cacheline addresses) to predict future memory accesses. 
These techniques either \rbc{completely} neglect a prefetcher’s undesirable effects \rbc{(e.g., memory bandwidth usage)} on the overall system, or \rbc{incorporate \rbc{system-level} feedback as \rbc{an afterthought} to a system-unaware prefetch algorithm}.
\rbc{We show that prior prefetchers often lose their performance benefit over a wide range of workloads and system \rbc{configurations} due to their inherent inability to take multiple different types of program context and system-level feedback information into account while prefetching.}
In this paper, we make a case for \rbc{designing a holistic prefetch algorithm that learns to prefetch using multiple \rbc{different types of} program context and system-level feedback \rbc{information inherent to its design}}.

To this end, we propose \emph{Pythia}, which formulates the prefetcher as a reinforcement learning agent.
For every demand request, Pythia observes multiple different types of program context information 
to \rbc{make} a prefetch decision. For every prefetch decision, Pythia receives a numerical reward that evaluates prefetch quality under the current memory bandwidth usage. Pythia uses this reward to reinforce the correlation between program context information and prefetch decision to generate highly accurate, timely, and system-aware prefetch requests in the future.
\rbc{Our} extensive evaluations using simulation and \rbc{hardware} synthesis show that Pythia outperforms \rbc{two} state-of-the-art prefetchers \rbc{(MLOP and Bingo)} by \rbc{$3.4$}\% and \rbc{$3.8$}\% in single-core, \rbc{$7.7$}\% and $9.6$\% in twelve-core, and \rbc{$16.9$}\% and \rbc{$20.2$}\% in bandwidth-constrained core configurations, while \rbc{incurring} only $1.03$\% area overhead \rbc{over} a desktop-class processor and no \rbc{software} changes in workloads.
The source code of Pythia can be \rbc{freely} downloaded from 
\url{https://github.com/CMU-SAFARI/Pythia}.


\end{abstract}

\maketitle
\thispagestyle{firstpage}

\section{Introduction} \label{sec:intro}

Prefetching is a well-studied speculation technique that predicts the addresses of long-latency memory requests and fetches the corresponding data from main memory to on-chip caches before the \rbc{program executing on the processor} demands it. 
\rbc{A program's repeated accesses over its data structures create patterns in its memory request addresses}.
A prefetcher tries to identify such memory access patterns from past memory requests to predict \rbc{the addresses of} future memory requests. To quickly identify a memory access pattern, a prefetcher typically uses some program context information to examine only \rbc{a subset} of memory requests. We call this program context a \emph{feature}. 
\rbc{The prefetcher associates a memory access pattern with a feature and generates prefetches following the same pattern \rbc{when} the feature reoccurs \rbc{during program execution}}.

\rbc{Past research has} proposed numerous prefetchers that consistently pushed the limits of prefetch coverage (i.e., the fraction of memory requests predicted by the prefetcher) and accuracy (i.e., the fraction of \rbc{prefetch} requests that are \rbc{actually} demanded by the program) by exploiting various program features, e.g., program counter (\texttt{PC}), cacheline address (\texttt{Address}), page offset of a cacheline (\texttt{Offset}), or \rbc{a simple} combination of such features using simple operations like concatenation (\texttt{+}) ~\cite{stride,streamer,baer2,stride_vector,jouppi_prefetch,ampm,fdp,footprint,sms,sms_mod,spp,vldp,sandbox,bop,dol,dspatch,bingo,mlop,ppf,ipcp}. 
For example, a PC-based stride prefetcher~\cite{stride,stride_vector,jouppi_prefetch} uses PC as the feature to learn the constant stride between two consecutive memory accesses \rbc{caused} by the same PC. VLDP~\cite{vldp} and SPP~\cite{spp} use a sequence of cacheline address deltas as the feature to predict the next cacheline \rbc{address} delta. Kumar and Wilkerson~\cite{footprint} use \texttt{PC+Address} of the first access \rbc{in} a \rbc{memory} region as the feature to predict the \rbc{spatial} memory access footprint \rbc{in} the entire memory region. SMS~\cite{sms} empirically finds \texttt{PC+Offset} of the first access \rbc{in a memory region} to be a better feature to predict the \rbc{memory access} footprint. Bingo~\cite{bingo} combines the features from~\cite{footprint} and SMS and uses \rbc{\texttt{PC+Address} and \texttt{PC+Offset}} \rbc{as its features}.

\rbc{Accurate and timely prefetch requests reduce} the long memory access latency \rbc{experienced by} the \rbc{processor}, \rbc{thereby} improving overall \rbc{system} performance. \rbc{However}, \rbc{speculative prefetch requests} \rbc{can} cause \rbc{undesirable} effects on the system (e.g., increased memory bandwidth consumption, cache pollution, \rbc{memory access interference,} etc.), which can reduce or negate the performance improvement gained by hiding memory access latency~\cite{fdp,ebrahimi2009coordinated}. \rbc{Thus, a good prefetcher aims to maximize its benefits while minimizing its undesirable effects on the system.}





\rbc{Even though there is a large number of prefetchers proposed in the literature, we observe 
three key shortcomings in almost every prior prefetcher design that significantly limits \rbc{its} performance benefits over a wide range of workloads and system configurations:
(1) \rbc{the use of \rbc{mainly} a single program feature for prefetch prediction}, (2) lack of inherent system awareness, and (3) \rbc{lack of ability to customize \rbc{the} prefetcher design to seamlessly adapt to a wide range of workload \rbc{and system configurations}}}.

\begin{figure}[!h]
\vspace{-0.8em}
\centering
\includegraphics[width=3.3in]{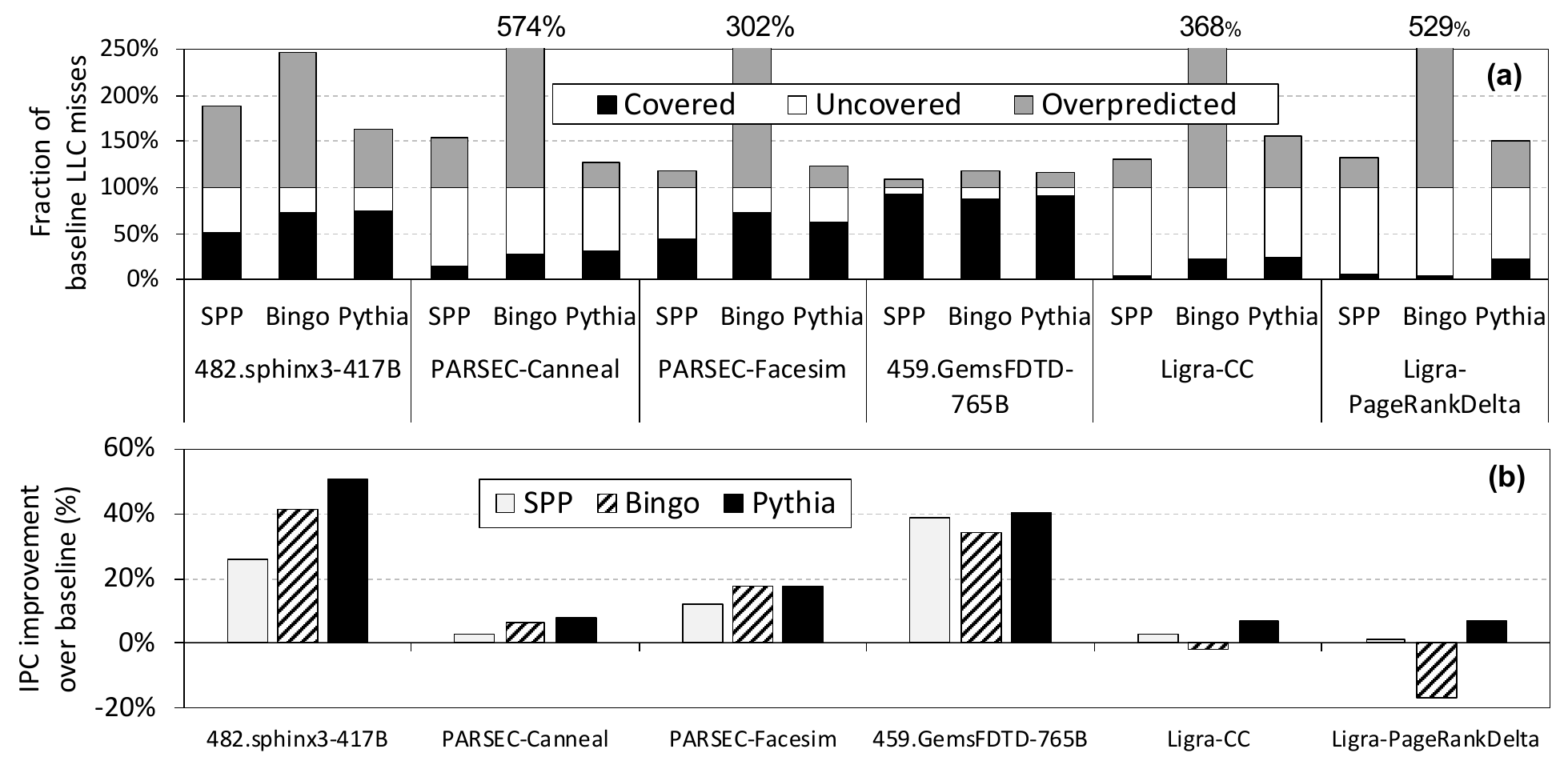}
\vspace{-0.5em}
\caption{Comparison of \rbc{(a) coverage, overprediction, and (b) performance of two recently-proposed prefetchers, SPP~\cite{spp} and Bingo~\cite{bingo}, and our new proposal, Pythia.}}
\label{fig:intro_perf_cov_acc}
\vspace{-1em}
\end{figure}

\begin{sloppypar}
\textbf{Single-feature prefetch prediction.} \rbc{Almost every prior prefetch-er relies on \emph{only one} program feature to correlate with the program memory access pattern and generate prefetch requests~\cite{stride,streamer,baer2,stride_vector,jouppi_prefetch,ampm,fdp,footprint,sms,sms_mod,spp,vldp,sandbox,bop,dol,dspatch,mlop,ppf,ipcp}. As a result, a prefetcher typically \rbc{provides} good \rbc{(or poor)} performance benefits in \rbc{mainly} those workloads where the correlation between the feature used by the prefetcher and program's memory access pattern is dominantly present \rbc{(or absent)}.}
\rbc{To demonstrate this, we show the coverage and overpredictions (i.e., \rbc{prefetched} memory requests that do \emph{not} get demanded by the processor) of two recently proposed prefetchers, SPP~\cite{spp} and Bingo~\cite{bingo}, and our new proposal Pythia (\cref{sec:design}) for six \rbc{example} workloads (\cref{sec:methodology} discusses our experimental methodology) in Fig.~\ref{fig:intro_perf_cov_acc}(a). Fig.~\ref{fig:intro_perf_cov_acc}(b) shows the performance of SPP, Bingo and Pythia on the same workloads.}
As we see in Fig.~\ref{fig:intro_perf_cov_acc}(a), Bingo provides higher prefetch coverage than SPP in \texttt{sphinx3}, \texttt{PARSEC-Canneal}, and \texttt{PARSEC-Facesim}, where the correlation exists between the first access \rbc{in} a \rbc{memory} region and the other accesses \rbc{in} the same region. \rbc{As a result, Bingo performs better than SPP in these workloads (Fig.~\ref{fig:intro_perf_cov_acc}(b)).} 
\rbc{In contrast}, for workloads like \texttt{GemsFDTD} that have regular access patterns within a physical page, SPP's \emph{sequence of deltas} feature \rbc{provides} better coverage and performance than Bingo.
\end{sloppypar}

\begin{sloppypar}
\textbf{Lack of inherent system awareness.}
\rbc{All prior prefetchers either completely neglect their undesirable effects on the system (e.g., memory bandwidth usage, cache pollution, memory access interference, system energy consumption, etc.)~\cite{stride,streamer,baer2,stride_vector,jouppi_prefetch,ampm,footprint,sms,sms_mod,spp,vldp,sandbox,bop,dol,bingo,mlop,ppf,ipcp} or incorporate system awareness as \rbc{an afterthought \rbc{(i.e., a separate control component)}} to the underlying system-unaware prefetch algorithm~\cite{fdp,ebrahimi2009coordinated,ebrahimi2009techniques,ebrahimi_paware,dspatch,bapi,pa_dram,mutlu2005,wflin,zhuang,wflin2,charney}. Due to the lack of inherent system awareness,}
a prefetcher often loses its performance gain in resource-constrained scenarios.
\rbc{For example, as shown in} Fig.~\ref{fig:intro_perf_cov_acc}(a), Bingo achieves similar \rbc{prefetch} coverage in \texttt{Ligra-CC} as compared to \texttt{PARSEC-Canneal}, while generating significantly lower overpredictions in \texttt{Ligra-CC} than \texttt{PARSEC-Canneal}.
However, 
Bingo loses performance in \texttt{Ligra-CC} by $1.9$\% compared to a no-prefetching baseline, whereas it improves performance by $6.4$\% in \texttt{PARSEC-Canneal} (Fig.~\ref{fig:intro_perf_cov_acc}(b)).
\rbc{
This contrasting outcome is due to Bingo's lack of awareness \rbc{of} the memory bandwidth usage.
\rbc{Without prefetching, }\texttt{Ligra-CC} consumes higher memory bandwidth than \texttt{PARSEC-Canneal}. As a result, 
\rbc{each overprediction}
made by Bingo in \texttt{Ligra-CC} wastes more precious \rbc{memory} bandwidth and is more detrimental to performance than 
\rbc{that in}
\texttt{PARSEC-Canneal}.}
\end{sloppypar}

\textbf{Lack of online prefetcher design customization.}
\rbc{The high design complexity of architecting a multi-feature, system-aware prefetcher has traditionally compelled architects to statically select only one program feature at design time. With every new prefetcher, architects design new rigid hardware structures to exploit the selected program feature.}
To exploit a new program feature for higher performance benefits, one must design a new prefetcher from scratch \rbc{and} extensively evaluate and verify \rbc{it both \rbc{in} pre-silicon and post-silicon realization}.
Due to the rigid design-time decisions, the hardware structures proposed by prior prefetchers cannot be customized \rbc{online} in silicon either to exploit any other program feature or to change the prefetcher’s objective (\rbc{e.g.}, to increase/decrease coverage, accuracy, or timeliness) \rbc{so that it can seamlessly adapt to} varying workloads and system configurations.

\vspace{0.3em}
\emph{\textbf{Our goal}} in this work is to design a single prefetching framework that (1) can holistically learn to prefetch using both \emph{multiple different types of program features} and \emph{system-level feedback} \rbc{information that is inherent to the design}, and (2) can be \emph{easily customized} in silicon via simple configuration registers to exploit 
\rbc{different types of program features and/or}
to change the objective of the prefetcher \rbc{(e.g., increasing/decreasing coverage, accuracy, or timeliness)} without any changes to the underlying hardware.

\vspace{0.3em}
\textbf{Key ideas.} To this end, we propose Pythia,\footnote{Pythia, according to Greek mythology, is the oracle of Delphi who is known for accurate prophecies~\cite{pythia}.} which formulates hardware prefetching as a reinforcement learning problem. Reinforcement learning (RL)~\cite{rl_bible,rlmc} is a machine learning paradigm that studies how an autonomous agent can learn to take optimal actions that maximizes a reward function by interacting with a stochastic environment.
We formulate Pythia as an RL-agent that autonomously learns to prefetch by interacting with the processor and the memory subsystem.
For every new demand request, Pythia extracts a set of program features.
It uses the set of features as \emph{state} information to take a prefetch \emph{action} based on its prior experience. For every prefetch action (including \rbc{\emph{not to prefetch}}), Pythia receives a numerical \emph{reward}
\rbc{which} evaluates the accuracy and timeliness of the prefetch action 
\rbc{given various system-level feedback \rbc{information}. While Pythia's framework is general enough to incorporate any type of system-level feedback \rbc{information} \rbc{into its decision making}, in this paper we demonstrate Pythia using \rbc{one} major system-level \rbc{information for prefetching}: memory bandwidth usage.} 
Pythia uses the reward received for a prefetch action to reinforce the \rbc{correlations} between various program features and the prefetch action and \rbc{learn from} experience \rbc{how} to generate accurate, timely, and system-aware prefetches in the future.
\rbc{The types of program feature used by Pythia and the reward level values can be easily customized in silicon via configuration registers.}

\textbf{Novelty and Benefits.}
Pythia's RL-based design approach requires an architect to only specify \emph{which} of the possible program features \emph{might} be useful to design a good prefetcher and \emph{what} performance goal the prefetcher should target, rather than spending time on designing and implementing a new \rbc{(likely rigid)} prefetch algorithm and accompanying rigid hardware that describes \emph{precisely how} the prefetcher should exploit the selected features to achieve that performance goal.
This \rbc{approach} provides two unique advantages over prior prefetching proposals.
First, using the RL framework, Pythia can holistically learn to prefetch using \rbc{both} \rbc{\emph{multiple program features}} and \emph{system-level feedback} \rbc{information} \rbc{inherent to its design}.
\rbc{Second, Pythia can be easily customized in silicon via simple configuration registers to exploit different types of program features and/or change the objective of the prefetcher. This gives Pythia the unique benefit of providing \rbc{even higher} performance \rbc{improvements} for a wide variety of workloads \rbc{and changing system configurations}, without any changes \rbc{to} the underlying hardware.}


\textbf{Results Summary.} We evaluate Pythia using a diverse set of memory-intensive workloads spanning \texttt{SPEC CPU2006}~\cite{spec2006}, \texttt{SPEC CPU2017}~\cite{spec2017}, \texttt{PARSEC 2.1}~\cite{parsec}, \texttt{Ligra}~\cite{ligra}, and \texttt{Cloudsuite}~\cite{cloudsuite} \rbc{benchmarks}.
\rbc{\rbc{We demonstrate four key results.} First, Pythia outperforms \rbc{two state-of-the-art prefetchers (MLOP~\cite{mlop} and Bingo~\cite{bingo}) by \rbc{$3.4$}\% and \rbc{$3.8$}\% in single-core and \rbc{$7.7$}\% and $9.6$\% in twelve-core configurations.}
This is because Pythia generates lower overpredictions, while simultaneously providing higher prefetch coverage than the prior prefetchers. 
Second, Pythia's performance benefits increase in bandwidth-constrained system configurations. For example, in a server-like configuration, where a core \rbc{can have} only $\frac{1}{16}\times$ \rbc{of the} bandwidth of a single-channel DDR4-$2400$~\cite{ddr4} DRAM controller, Pythia outperforms MLOP and Bingo by $16.9$\% and $20.2$\%.
Third, Pythia can be customized further via \rbc{simple} configuration registers \rbc{to target} workload suites to provide even higher performance benefits.
We demonstrate that by simply changing the numerical rewards, Pythia provides up to $7.8$\% ($1.9$\% on average) more performance improvement across all \texttt{Ligra} \rbc{graph processing} workloads over the \rbcb{basic} Pythia configuration.
}
Fourth, Pythia's performance benefits come with \rbc{only} modest area and power \rbc{overheads}. 
\rbc{Our functionally-verified hardware synthesis for Pythia shows that}
Pythia only incurs an area and power overhead of $1.03$\% and $0.37$\% over a $4$-core desktop-class processor.
\vspace{0.5em}
\noindent We make the following contributions in this paper:
\begin{itemize}
    \item \rbc{We observe three key shortcomings in prior prefetchers that significantly limits their performance benefits:
    (1) \rbc{the use of only a single program feature for prefetch prediction, (2) lack of inherent system awareness, and (3) lack of ability to customize \rbc{the} prefetcher design to seamlessly adapt to a wide range of workloads \rbc{and system configurations}}.
    }
    \item \rbc{We introduce a new prefetcher called Pythia}. Pythia formulates the prefetcher as a reinforcement learning (RL) agent, which takes adaptive prefetch decisions \rbc{by autonomously learning using both multiple program features and system-level feedback information \rbc{inherent to its design}} (\cref{sec:key_idea_formulation}).
    \item \rbc{We provide a low-overhead, practical implementation of Pythia's \rbc{RL-based} algorithm in hardware, which uses no more complex structures than simple tables (\cref{sec:ke_design}). \rbc{This design can potentially be used for other hardware structures that can benefit from RL principles.}}
    \item By extensive evaluation, we show that Pythia outperforms prior state-of-the-art prefetchers over a \rbc{wide} variety of workloads in a wide range of system configurations.
    \item \rbc{We open source Pythia and all the workload traces used for performance modeling in our GitHub repository: \url{https://github.com/CMU-SAFARI/Pythia}.}
\end{itemize}
\section{Background}\label{sec:background}

\rbc{We first} briefly review the basics of reinforcement learning~\cite{rl_bible, rlmc}. We then \rbc{describe} why reinforcement learning is a good \rbc{framework} for designing a hardware prefetcher \rbc{that fits our goals}.

\subsection{Reinforcement Learning}\label{sec:background_rl}

\begin{figure}[!h]
\vspace{-0.5em}
\centering
\includegraphics[scale=0.3]{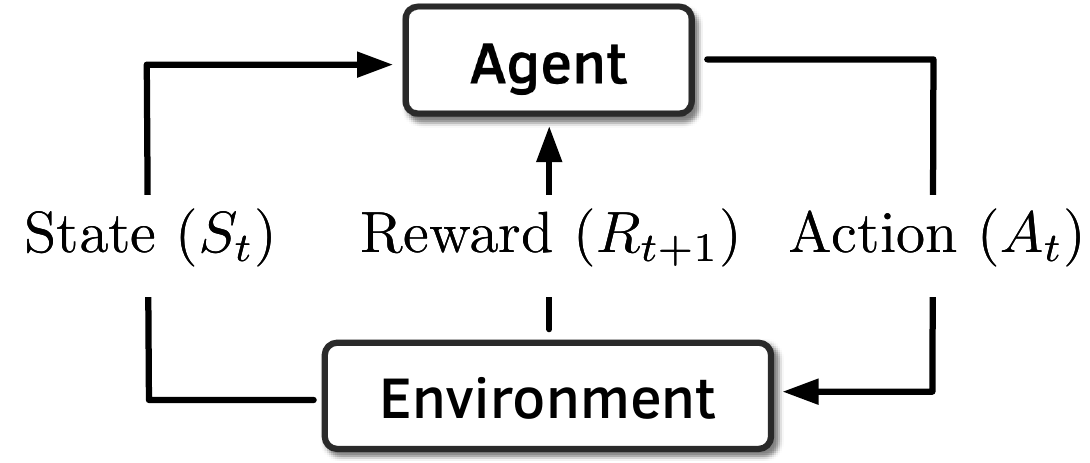}
\vspace{-1em}
\caption{Interaction between an agent and the environment in a reinforcement learning system.}
\vspace{-1em}
\label{fig:rl_basics}
\end{figure}

Reinforcement learning (RL)~\cite{rl_bible,rlmc}, in its simplest form, is the algorithmic approach to learn how to take \rbc{an} \emph{action} in a given \emph{situation} to maximize a numerical \emph{reward} signal. 
A typical RL system \rbc{comprises} of two main components: \emph{the agent} and \emph{the environment}, as shown in Fig.~\ref{fig:rl_basics}. The agent is the entity \rbc{that takes actions}.
\rbc{the agent resides in the environment and interacts with it in discrete timesteps.}
At each timestep $t$, the agent observes the current \textbf{\emph{state}} of the environment \textbf{$S_t$} and takes \textbf{\emph{action}} \textbf{$A_t$}. Upon receiving the action, the environment transitions to a new state $\mathbf{S_{t+1}}$, and emits an immediate \textbf{\emph{reward}} $R_{t+1}$, which is \rbc{immediately or} later \rbc{delivered} to the agent. The reward scheme encapsulates \rbc{the} agent's objective and drives the agent \rbc{towards taking} optimal actions.

\rbc{The \textbf{\emph{policy}} of the agent dictates it to take a certain action in a given state. \emph{The agent's goal is to find the optimal policy that maximizes the cumulative reward collected from the environment over time.}}
The expected cumulative reward by taking an action $A$ in a given state $S$ is defined as the \textbf{\emph{Q-value}} of the state-action pair (denoted as $Q(S,A)$).
\rbc{At} every timestep $t$, the agent iteratively optimizes its policy in two steps: (1) the agent updates the Q-value of a state-action pair using \rbc{the reward} collected in the current timestep, and (2) the agent \rbc{optimizes its current policy} using the newly updated Q-value.

\textbf{Updating \rbc{Q-values}.} If at a given timestep $t$, the agent observes a state $S_t$, takes an action $A_t$, while the environment transitions to a new state $S_{t+1}$ and emits a reward $R_{t+1}$ and the agent takes action $A_{t+1}$ in the new state, the Q-value of the old state-action pair $Q(S_t, A_t)$ is iteratively optimized using \rbc{the} SARSA~\cite{sarsa, rl_bible} \rbc{algorithm,} as shown in Eqn.~\eqref{eq:sarsa}\rbc{:}

\begin{equation}\label{eq:sarsa}
\normalsize
\begin{aligned}
Q\left(S_t, A_t\right) & \gets Q\left(S_t, A_t\right)\\
&+ \alpha\left[R_{t+1}+\gamma Q\left(S_{t+1}, A_{t+1}\right) - Q\left(S_t, A_t\right)\right]
\end{aligned}
\end{equation} 

$\alpha$ is the \emph{learning rate} parameter \rbc{that} controls the convergence rate of Q-values.
$\gamma$ is the \emph{discount factor}, \rbc{which is used} to assign more weight to the immediate reward received by the agent at any given timestep than to the delayed future rewards. A $\gamma$ value closer to 1 gives a ``far-sighted" planning capability to the agent, i.e., the agent can trade off a low immediate reward to gain higher rewards in the future. This is particularly useful in creating an autonomous agent that can anticipate the \rbc{long-term} \rbc{effects} of taking an action to optimize its policy \rbc{that gets closer to optimal over time}.

\textbf{\rbc{Optimizing} policy.} To find a policy that maximizes the cumulative reward collected over time, a purely-greedy agent always exploits the action $A$ in a given state $S$ that provides the highest Q-value $Q(S,A)$. However, greedy exploitation can leave the state-action space under-explored.
Thus, in order to strike a balance between exploration and exploitation, an $\epsilon$-greedy agent \emph{stochastically} takes a random action with a low probability of $\epsilon$ (called \emph{exploration rate}); otherwise, it selects the action that provides the highest Q-value~\cite{rl_bible}. 

In short, the Q-value serves as the foundational cornerstone of reinforcement learning.
By iteratively learning Q-values of state-action pairs, an RL-agent \rbc{continuously optimizes its policy} to take actions \rbc{that get closer to optimal over time}. 

\vspace{-5pt}
\subsection{Why \rbc{is} RL a Good Fit for Prefetching?} \label{sec:background_rl_prefetching}
The RL framework \rbc{has} been recently successfully demonstrated to solve complex problems like mastering human-like control \rbc{on} Atari~\cite{deepmind_atari} and Go~\cite{alpha_go, alpha_zero}.
\rbc{We argue that the RL framework is an inherent fit to model a hardware prefetcher for three key reasons.}

    \textbf{Adaptive learning in \rbc{a complex state space}.}
    As we state in \cref{sec:intro}, the \rbc{benefits} of a prefetcher not only depends on its coverage and accuracy but also on its \rbc{undesirable effects} on the system, like \rbc{memory bandwidth usage}.
    In other words, \emph{it is not sufficient for a prefetcher only to make highly accurate predictions}. Instead, a prefetcher should be \emph{performance-driven}. 
    A prefetcher should have the capability to adaptively trade-off coverage for higher accuracy (and vice-versa) depending on its impact on the overall system to provide a robust performance improvement with varying \rbc{workloads and system configurations}. This adaptive and performance-driven nature of prefetching \rbc{in a complex state space} makes RL a good fit for modeling a prefetcher \rbc{as an autonomous agent that learns to prefetch by interacting with the system.}
    
    \textbf{Online learning.}
    \rbc{An RL agent \emph{does not} require an expensive offline training phase. Instead, it can \emph{continuously} learn \emph{online} by iteratively optimizing its policy using the rewards received from the environment. A hardware prefetcher, similar to an RL agent, also needs to continuously learn from the changing workload behavior and system \rbc{conditions} to provide consistent performance benefits. The online learning requirement of prefetching makes RL an inherent fit to model a hardware prefetcher.}

    \begin{sloppypar}
    \textbf{Ease of implementation.}
    Prior works have evaluated many sophisticated \rbc{machine} learning models like simple neural \rbc{networks}~\cite{peled2018neural}, \rbc{LSTMs}~\cite{hashemi2018learning, shineural}, and Graph Neural \rbc{Networks} (GNNs)~\cite{shi2019learning} as \rbc{models for} hardware \rbc{prefetching}. \rbc{Even though} these techniques show encouraging results in accurately predicting memory accesses, they fall short \rbc{especially} in two major aspects. First, these models' \rbc{sizes} often exceed even the \rbc{largest} \rbc{caches in traditional processors}~\cite{peled2018neural,hashemi2018learning,shineural,shi2019learning}, making them impractical \rbc{(or at best very difficult) to implement}. 
    Second, due to the vast amount of computation \rbc{they require for inference}, these models’ inference latency is much higher than an acceptable latency of a prefetcher at any cache level.
    On the other hand, we can efficiently implement an RL-based model, as we \rbc{demonstrate} in this paper \rbc{(\cref{sec:design})}, that can \emph{quickly} make predictions and can be \rbc{relatively} easily adopted in a real processor. 
    \end{sloppypar}
    
\section{Pythia: Key Idea}
\begin{figure}[!h]
\vspace{-0.5em}
\centering
\includegraphics[scale=0.3]{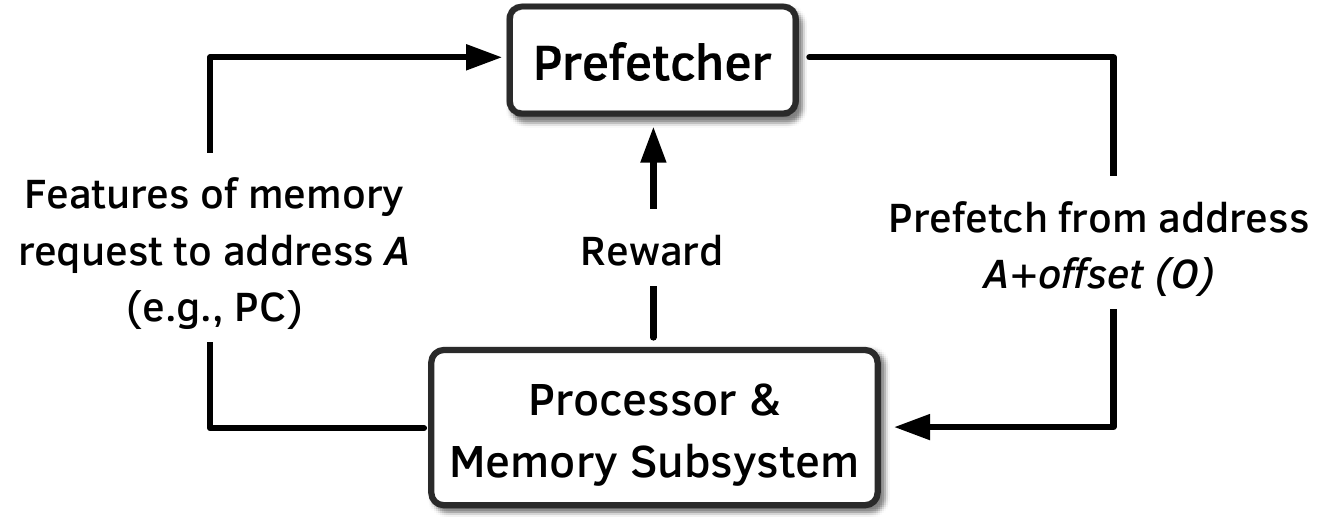}
\vspace{-1em}
\caption{Formulating \rbc{the} prefetcher as an RL-agent.}
\label{fig:rl_as_prefetcher}
\vspace{-1.2em}
\end{figure}

In this work, we formulate prefetching as a reinforcement learning problem, as shown in Fig.~\ref{fig:rl_as_prefetcher}. Specifically, we formulate Pythia as an RL-agent that learns to make accurate, timely, and system-aware prefetch decisions by interacting with the environment, i.e., the processor and the memory subsystem.
Each timestep corresponds to a new demand request seen by Pythia. With every new demand request, Pythia observes the state of the processor and the memory subsystem and takes a prefetch action.
\rbc{For every prefetch action (including \rbc{\emph{not to prefetch}}), Pythia receives a numerical reward \rbc{that} evaluates the accuracy and timeliness of the prefetch action \rbc{taking into account} various system-level feedback information. 
\emph{Pythia's goal is to find the optimal prefetching policy that would maximize the number of accurate and timely prefetch requests, \rbc{taking} system-level feedback information \rbc{into account}}. While Pythia's framework is general enough to incorporate any type of system-level feedback \rbc{into its decision making}, in this paper we demonstrate Pythia using \rbc{\emph{memory bandwidth usage}} as the system-level feedback information.}

\subsection{Formulation of the RL-based Prefetcher}\label{sec:key_idea_formulation}
We formally define the three pillars of our RL-based prefetcher: the state \rbc{space}, \rbc{the} actions, and the \rbc{reward} scheme.

\textbf{State.} 
We define the state as a $k$-dimensional vector of program features. 
\begin{equation}\label{eq:state}
    S \equiv \{ \phi_{S}^{1}, \phi_{S}^{2}, \ldots, \phi_{S}^{k} \}
\end{equation}
Each program feature is composed of \rbc{at most} two components: (1) program control-flow component, and (2) program data-flow component. \rbc{The} control-flow component is further made \rbc{up} of simple information like \rbc{load-PC (i.e., the PC of a load instruction)} or \rbc{branch-PC (i.e., the PC of a branch instruction that immediately precedes a load instruction)}, and a history that denotes whether this information is extracted only from the current demand request or a series of past demand requests.  
Similarly, the data-flow component is made \rbc{up} of simple information like cacheline address, \rbc{physical} page \rbc{number}, page offset, cacheline delta, and its corresponding history. 
Table~\ref{table:features} shows some example program features. 
\rbc{Although} Pythia can \rbc{theoretically} learn to prefetch \rbc{using} \emph{any} number of such program features, we fix the state-vector dimension (i.e., $k$) at design time \rbc{given a limited storage budget in hardware}. However, the exact selection of $k$ program features out of all possible program features is configurable \rbc{online} using simple configuration registers. 
In \cref{sec:tuning_feature_selection}, we \rbc{provide} an \emph{automated feature selection method} to find a vector of program features \rbc{to be used at design time}.


\begin{table}[htbp]
  \centering
  \footnotesize
  \caption{Example program features}
  \vspace{-0.5em}
    \begin{tabular}{m{8.8em}C{3em}C{4em}C{6.7em}C{4em}}
    \toprule
    \multicolumn{1}{l}{\multirow{2}[4]{*}{\textbf{Feature}}} & \multicolumn{2}{c}{\textbf{Control-flow}} & \multicolumn{2}{c}{\textbf{Data-flow}} \\
    \cmidrule{2-5} & \textbf{Info.} & \textbf{History} & \textbf{Info.} & \textbf{History} \\
    \midrule
    Last 3-PCs & PC & last 3 & \xmark  & \xmark \\
    Last 4-deltas & \xmark  & \xmark  & Cacheline delta & last 4 \\
    PC+Delta & PC & current & Cacheline delta & current \\
    Last 4-PCs+Page \rbc{no.} & PC & last 4 & Page \rbc{no.} & current \\ 
    \bottomrule
    \end{tabular}%
  \label{table:features}%
  \vspace{-2em}
\end{table}%


\textbf{Action.} 
We define the action of the RL-agent as selecting a \rbc{\emph{prefetch offset}} (i.e., \rbc{a} delta, \rbc{"O" in Fig.~\ref{fig:rl_as_prefetcher}}, between the predicted and the \rbc{demanded} cacheline address) from a set of candidate prefetch offsets.
As every post-L1\rbc{-cache} prefetcher generates prefetch requests within a physical page~\cite{ampm,fdp,footprint,sms,sms_mod,spp,vldp,sandbox,bop,dol,dspatch,bingo,mlop,ppf,ipcp}, the list of prefetch offsets only contains values in the range of $[-63,63]$ for a system with a traditionally-sized $4$KB page and $64$B cacheline. 
Using prefetch offsets as \rbc{actions} (instead of full cacheline addresses)
drastically reduces the action space size. We further reduce the action space size by fine tuning, as described in \cref{sec:tuning_action_selection}.
\rbc{A prefetch offset of} zero means no \rbc{prefetch is generated}.

\textbf{Reward.}
The reward structure defines the prefetcher's objective.
We define five different reward levels as follows.
\begin{itemize}
    \item \textbf{\emph{Accurate and timely}} ($\mathcal{R}_{AT}$). This reward is assigned to an action whose corresponding prefetch address gets demanded \emph{after} the prefetch fill.
    \item \textbf{\emph{Accurate but \rbc{late}}} ($\mathcal{R}_{AL}$). This reward is assigned to an action whose corresponding prefetch address gets demanded \emph{before} the prefetch fill.
    \item \textbf{\emph{Loss of coverage}} ($\mathcal{R}_{CL}$). This reward is assigned to an action whose corresponding prefetch address \rbc{is to a different physical page than the demand access that led to the prefetch.}
    \item \textbf{\emph{Inaccurate}} ($\mathcal{R}_{IN}$). This reward is assigned to an action whose corresponding prefetch address does \emph{not} get demanded in a temporal window. The reward is classified into two sub-levels: inaccurate given low bandwidth \rbc{usage} ($\mathcal{R}_{IN}^{L}$) and inaccurate given high bandwidth \rbc{usage} ($\mathcal{R}_{IN}^{H}$).
    \item \textbf{\emph{No-prefetch}} ($\mathcal{R}_{NP}$). This reward is assigned when Pythia decides not to prefetch. This reward level is also classified into two sub-levels: no-prefetch given low bandwidth \rbc{usage} ($\mathcal{R}_{NP}^{L}$) and no-prefetch given high bandwidth \rbc{usage} ($\mathcal{R}_{NP}^{H}$).
\end{itemize}

\rbc{By increasing (decreasing) a reward level value, we reinforce (deter) Pythia to collect such rewards from the environment in the future.}
$\mathcal{R}_{AT}$ and $\mathcal{R}_{AL}$ \rbc{are} used to guide Pythia to generate more accurate and timely prefetch requests.
$\mathcal{R}_{CL}$ is used to guide Pythia to generate prefetches within the physical page of the triggering demand request.
$\mathcal{R}_{IN}$ and $\mathcal{R}_{NP}$ are used to define Pythia's prefetching strategy with respect to memory bandwidth usage feedback.
In \cref{sec:tuning_reward_hyp_selection}, we \rbc{provide} an \emph{automated method} to configure the reward values.
The reward values can be easily customized further for target workload suites to extract higher performance gains (\cref{sec:eval_pythia_custom}).
\section{Pythia: Design} \label{sec:design}

\begin{figure}[!h]
\vspace{-1.2em}
\centering
\includegraphics[width=3.3in]{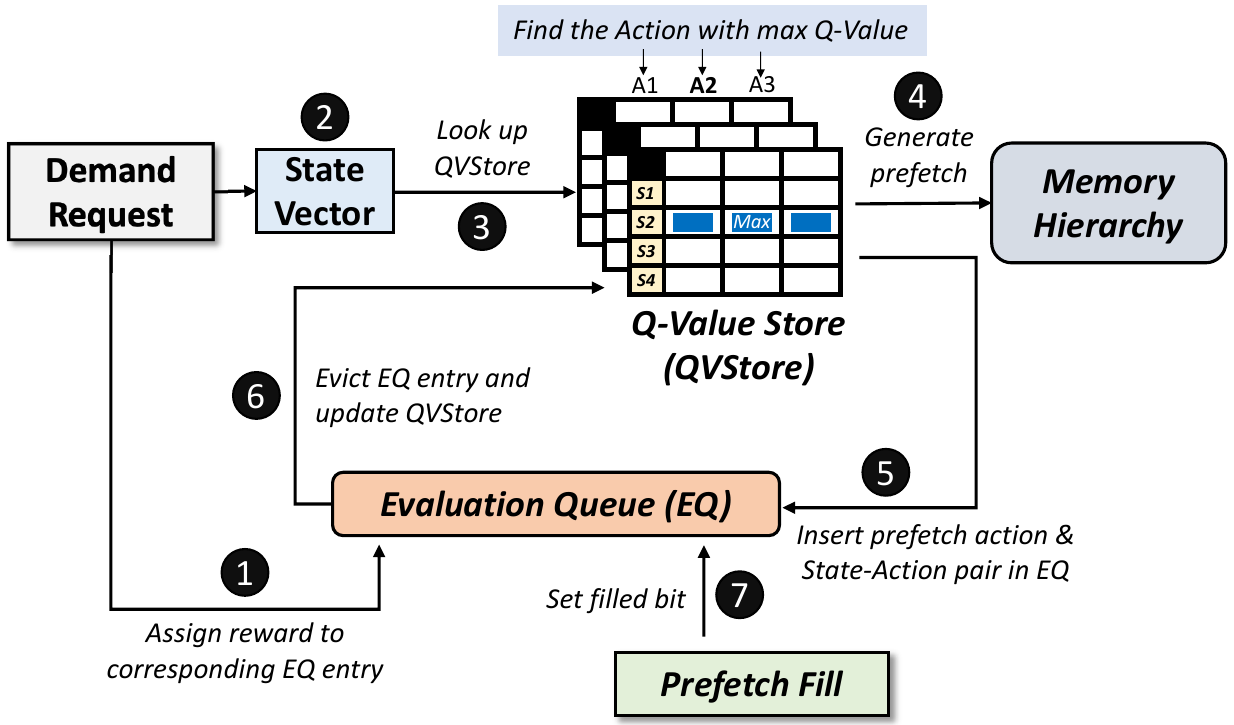}
\vspace{-1em}
\caption{Overview of Pythia.}
\vspace{-2em}
\label{fig:overview}
\end{figure}

Fig.~\ref{fig:overview} shows a \rbc{high-level} \rbc{overview} of Pythia. 
Pythia is mainly comprised of two hardware structures: \emph{Q-Value Store} (QVStore) and \emph{Evaluation Queue} (EQ).
The purpose of QVStore is to \rbc{record} Q-values for all state-action pairs that are \rbc{observed} by Pythia.
The purpose of EQ is to maintain a first-in-first-out list of Pythia's recently-taken actions.\footnote{Pythia keeps track of recently-taken actions because it cannot always \emph{immediately} assign a reward to an action, as the \rbc{usefulness} of the generated prefetch request (i.e., \rbc{if and when} the prefetched address is demanded by the processor) is not immediately known while the action is being taken.
During EQ residency, \rbc{if the address of a demand request matches with the prefetch address stored in an EQ entry}, the corresponding action is considered to \rbc{have generated a useful prefetch request}. 
}
\rbc{Every EQ entry holds three \rbc{pieces of} information: \rbc{(1)} the \rbc{taken} action, \rbc{(2)} the prefetch address generated for the corresponding action, and \rbc{(3)} a \emph{filled} bit. A set filled bit indicates that the prefetch request has been filled into the cache.}

For every new demand request, Pythia first checks the EQ with the demanded memory address (\circled{1}). \rbc{If the address is present in the EQ (i.e., Pythia has issued a prefetch request for this address in the past)},
\rbc{it signifies that the prefetch action corresponding to the EQ entry \rbc{has generated a useful prefetch request}. As such, Pythia assigns a reward (either $\mathcal{R}_{AT}$ or $\mathcal{R}_{AL}$) to the EQ entry, based on whether \rbc{or not} the EQ entry's filled bit is set.}
\rbc{Next, Pythia extracts the state-vector from the attributes of the demand request (e.g., PC, address, cacheline delta, etc.) (\circled{2}) and looks up QVStore to find the action with the maximum Q-value for the given state-vector (\circled{3}).}
\rbc{Pythia selects the action with the maximum Q-value to generate prefetch request and issues the request to the memory hierarchy (\circled{4}).}
\rbc{At the same time, Pythia inserts the selected prefetch action, its corresponding prefetched memory address, and the state-vector into EQ (\circled{5}).}
Note that, a \emph{no-prefetch} action or an action that prefetches an address beyond the current physical page
is also inserted into EQ. The reward for such \rbc{an} action is instantaneously assigned to the EQ entry.
\rbc{When an EQ entry gets evicted,} the state-action pair and the reward stored in the evicted EQ entry are used to update the Q-value in the QVStore (\circled{6}).
For every prefetch fill in cache, Pythia looks up EQ with the prefetch address \rbc{and sets the \emph{filled} bit in the matching EQ entry indicating that the prefetch request has been filled \rbc{into the} cache (\circled{7}).}
\rbc{Pythia uses this filled bit in \circled{1} to classify actions that generated timely or late prefetches.}\footnote{In this paper, we define prefetch timeliness as a binary value due to its measurement simplicity. One can easily make the definition non-binary by storing three timestamps per EQ entry: (1) when the prefetch is issued ($t_{issue}$), (2) when the prefetch is filled ($t_{fill}$), and (3) when a demand is generated for the same prefetched address ($t_{demand}$).}

\begin{algorithm*}[h]
  \footnotesize
  \caption{Pythia's \rbc{reinforcement learning} based prefetching algorithm}\label{algo:Pythia}
  \begin{algorithmic}[1]
  \Procedure{Initialize}{}
    \State initialize QVStore: $Q(S,A)\gets \frac{1}{1-\gamma}$ 
    \State clear EQ
  \EndProcedure
  \State
  \Procedure{Train\_and\_Predict}{Addr} \hspace{32pt} \textit{/* Called for every demand request */}

    \State $entry \gets search\_EQ(Addr)$ \hspace{46pt} \textit{/* For a demand request to $Addr$, search EQ with the demand address */}
    \If{entry is valid}
        \If {$entry.filled == true$}
            \State $entry.reward \gets \mathcal{R}_{AT}$ \hspace{40pt} \textit{/* If the filled bit is set, i.e., the demand access came \emph{after} the prefetch fill, assign reward $\mathcal{R}_{AT}$ */}
        \Else
            \State $entry.reward \gets \mathcal{R}_{AL}$ \hspace{40pt} \textit{/* Otherwise, assign $\mathcal{R}_{AL}$ */}
        \EndIf
    \EndIf
    \State $S\gets get\_state()$ \hspace{80pt} \textit{/* Extract the state-vector from the attributes of current demand request */}
    \If{$rand()\le \epsilon$}
        \State $action\gets get\_random\_action()$ \hspace{20pt} \textit{/* Select a random action with a low probability $\epsilon$ to explore the state-action space */}
    \Else
        \State $action\gets \argmax_a Q(S,a)$ \hspace{38pt} \textit{/* Otherwise, select the action with the highest Q-value */}
    \EndIf
    \State $prefetch(Addr+Offset[action])$ \hspace{20pt} \textit{/* Add the selected prefetch offset to the current demand address to generate prefetch address */}
    \State $entry \gets create\_EQ\_entry(S,action,Addr+Offset[action])$ \hspace{1.5em} \textit{/* Create new EQ entry using the current state-vector, the selected action, and the prefetch address */}
    \If {no prefetch action}
        \State $entry.reward \gets \mathcal{R}_{NP}^{H}$ or $\mathcal{R}_{NP}^{L}$ \hspace{23pt} \textit{/* In case of no-prefetch action, immediately assign reward ${R}_{NP}^{H}$ or ${R}_{NP}^{L}$ based on current memory bandwidth usage */}
    \ElsIf{out-of-page prefetch}
        \State $entry.reward \gets \mathcal{R}_{CL}$ \hspace{50pt} \textit{/* In case of out-of-page prefetch action, immediately assign reward ${R}_{CL}$ */}
    \EndIf
    \State $dq\_entry\gets insert\_EQ(entry)$ \hspace{25pt} \textit{/* Insert the entry. Get the evicted EQ entry. */}
    \If{$has\_reward(dq\_entry) == false$}
        \State $dq\_entry.reward \gets \mathcal{R}_{IN}^{H}$ or $\mathcal{R}_{IN}^{L}$ \hspace{2em} \textit{/* If the evicted entry does not have a reward yet, assign the reward $\mathcal{R}_{IN}^{H}$ or $\mathcal{R}_{IN}^{L}$ based on current memory bandwidth usage */}
    \EndIf
    \State $R \gets  dq\_entry.reward$ \hspace{116pt} \textit{/* Get the reward stored in the evicted entry */}
    \State $S_1\gets dq\_entry.state;\ \ A_1\gets dq\_entry.action$ \hspace{42pt} \textit{/* Get the state-vector and the action from the evicted EQ entry */}
    \State $S_2\gets EQ.head.state;\ \ A_2\gets EQ.head.action$ \hspace{46pt} \textit{/* Get the state-vector and the action from the entry at the head of the EQ */}
    \State $Q(S_1, A_1)\gets Q(S_1, A_1)+\alpha[R+\gamma Q(S_2, A_2)-Q(S_1, A_1)]$ \hspace{2em} \textit{/* Perform the SARSA update */}
  \EndProcedure
  \State
  \Procedure{Prefetch\_Fill}{Addr}
    \State $search\_and\_mark\_EQ(Addr,\ \texttt{FILLED})$ \hspace{2em} \textit{/* For every prefetch fill, search the address in EQ and mark the corresponding EQ entry as filled */}
  \EndProcedure
  \end{algorithmic}
\end{algorithm*}


\subsection{RL-based Prefetching Algorithm}
Algorithm~\ref{algo:Pythia} shows Pythia's RL-based prefetching algorithm. \rbc{Initially,} all entries in QVStore are
reset to the highest possible Q-value ($\frac{1}{1-\gamma}$) and the EQ is cleared (lines 2-3). 
For every demand \rbc{request} to \rbc{a} cacheline address $Addr$, Pythia searches for $Addr$ in EQ (line $6$). If a matching entry is found, Pythia \rbc{assigns a reward (either $\mathcal{R}_{AT}$ or $\mathcal{R}_{AL}$) based on the \emph{filled} bit in the EQ entry (lines 8-11)}. Pythia then extracts the state-vector to \emph{stochastically} select a prefetching action (Sec.~\ref{sec:background}) that provides the highest Q-value (lines 13-16). \rbc{Pythia uses the selected action to generate the prefetch request (line 17) and creates a new EQ entry with the current state-vector, the selected action, and its corresponding prefetched address (line 18).}
\rbc{In case of a no-prefetch action, or an action that prefetches beyond the current physical page, Pythia immediately assigns the reward to the newly-created EQ entry (lines 19-22).}
\rbc{The EQ entry is then inserted, which evicts an entry from EQ. If the evicted EQ entry does not already have a reward assigned (\rbc{indicating that} the corresponding prefetch address is \emph{not} demanded \rbc{by} the processor \rbc{so far}), Pythia assigns the reward $\mathcal{R}_{IN}^{H}$ or $\mathcal{R}_{IN}^{L}$ based on the current memory bandwidth usage (lines 25).}
\rbc{Finally, the Q-value of the evicted state-action pair is updated \rbc{via} the SARSA algorithm (Sec.~\ref{sec:background}), using the reward stored in the evicted EQ entry and the Q-value of the state-action pair in the head of the EQ-entry (lines 26-29).}

\subsection{Detailed Design of Pythia}
\rbc{We describe the organization of QVStore \rbc{(\cref{sec:ke_design})}, how Pythia searches QVStore to get the action with the maximum Q-value for a given state-vector (\circled{3}) \rbc{(\cref{sec:design_config_pipeline})}, how Pythia assigns rewards to each taken action and how it updates Q-values (\circled{6}) \rbc{(\cref{sec:ke_update})}.
}

\subsubsection{\textbf{Organization of QVStore}} \label{sec:ke_design}
\rbc{The purpose of QVStore is to record Q-values for all state-action pairs that Pythia observes.}
Unlike prior real-world applications of RL~\cite{deepmind_atari,alpha_go,alpha_zero}, which use deep neural networks to \emph{approximately} \rbc{store} Q-values of every state-action pair, we propose a \rbc{new}, table-based, hierarchical QVStore organization that is \rbc{custom-designed to} our RL-agent.
 
\begin{figure}[!h]
\vspace{-0.5em}
\centering
\includegraphics[width=3.3in]{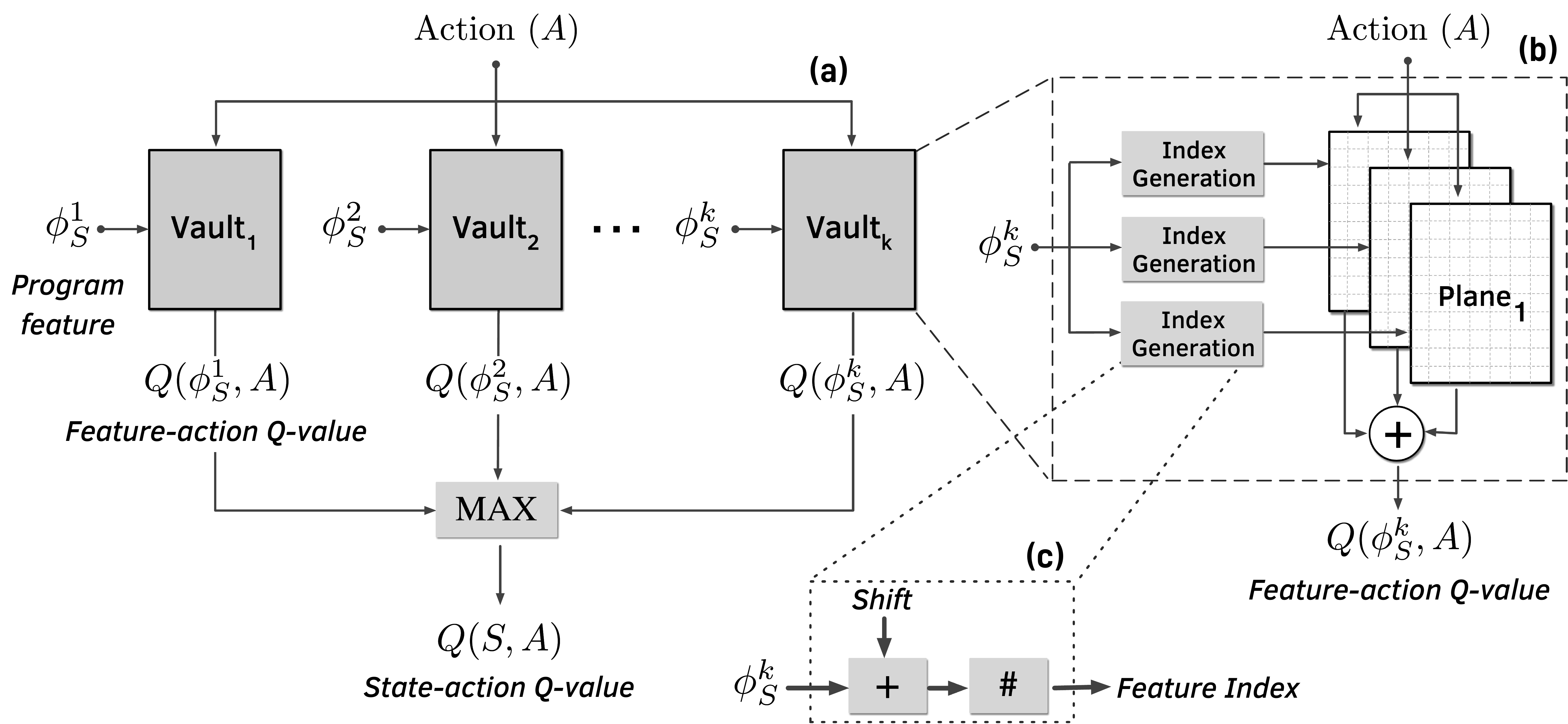}
\vspace{-0.5em}
\caption{(a) The QVStore is comprised of multiple vaults. (b) Each vault is comprised of multiple planes. (c) Index generation from feature value.}
\label{fig:ke_design}
\vspace{-1em}
\end{figure}

\rbc{Fig.~\ref{fig:ke_design}(a) shows the high-level organization of QVStore and how the Q-value is retrieved from QVStore for a given state $S$ (which is a k-dimensional vector of program features, $\{\phi_{S}^{1}, \phi_{S}^{2}, \ldots, \phi_{S}^{k}\}$) and an action $A$. As the state space grows rapidly with the state-vector dimension ($k$) and the bits used to represent each feature, we employ a hierarchical organization for QVStore. 
We organize QVStore in $k$ partitions, each \rbc{of which} we call a \emph{vault}. Each vault corresponds to one constituent feature of the state-vector and records the Q-values for the feature-action pair, $Q(\phi_{S}^{i},A)$.
}
During the Q-value retrieval for a given state-action pair $Q(S,A)$, Pythia queries each vault in parallel to retrieve the Q-values of constituent feature-action pairs $Q(\phi_{S}^{i},A)$. The final Q-value of the state-action pair $Q(S,A)$ is computed as the \emph{maximum} of all constituent feature-action Q-values, \rbc{as} Eqn.~\ref{eq:q_value} shows).
\rbc{The maximum operation ensures that the state-action Q-value is driven by the constituent feature of the state-vector that has the highest feature-action Q-value.}
\rbc{The vault organization enables QVStore to efficiently scale up \rbc{to} higher state-vector \rbc{dimensions: one} can increase the state-vector dimension by simply adding a new vault to the QVStore.}
\begin{equation} \label{eq:q_value}
    Q(S, A) = \max_{i \in (1, k)} Q(\phi_{S}^{i}, A)
\end{equation}

\rbc{\rbc{Fig.~\ref{fig:ke_design}(a) shows the organization of QVStore as a collection of multiple vaults.} 
The purpose of a vault is to record Q-values of all feature-action pairs that Pythia observes for a specific feature type.} A vault can be conceptually visualized as a monolithic two-dimensional table (as shown in Fig.~\ref{fig:ke_design}(a)), indexed by the feature and action values, that stores Q-value for every feature-action pair. 
\rbc{However, the key challenge in \rbc{implementing} vault as a monolithic table}
is that \rbc{the size of the table increases exponentially with a} linear increase in the number of bits used to represent the feature.
This not only makes the monolithic table \rbc{organization} impractical for implementation but also increases the \rbc{design} complexity to satisfy \rbc{its} latency and power requirements. 

One way to address this challenge is \rbc{to quantize} the feature space into \rbc{a} small number of tiles.
\rbc{Even though} feature space quantization can achieve a drastic reduction in the monolithic table size, 
\rbc{it requires a compromise between the resolution of a feature value and the generalization of feature values.}
We draw inspiration from \emph{tile coding}~\cite{rl_bible, cmac, rlmc} to strike a balance between resolution and generalization.
\rbc{Tile coding uses} \emph{multiple overlapping} hash functions to quantize a feature value into smaller \emph{tiles}. The quantization achieves generalization of similar feature values, whereas multiple hash functions increase resolution to represent a feature value.

We leverage the idea of tile coding to organize a vault as a collection of $N$ small two-dimensional tables, each \rbc{of which} we call a \emph{plane}. 
Each plane entry stores a \emph{partial} Q-value of a feature-action pair.\footnote{\rbc{Our application of tile coding is similar to that used in \rbc{the} self-optimizing memory controller (RLMC)~\cite{rlmc}. The key difference is that RLMC uses a hybrid combination of feature and action values to index \emph{single-dimensional} planes, whereas Pythia uses feature and action values \rbc{\emph{separately}} to index \emph{two-dimensional} planes.}} As \rbc{Fig.~\ref{fig:ke_design}(c)} shows, \rbc{to retrieve a feature-action Q-value $Q(\phi_{S}^{i},A)$,} the given feature is first shifted by a shifting constant \rbc{(which is randomly selected at design time)}, followed by a hashing to get the feature index for the given plane. This feature index, along with the action index, is used to \rbc{retrieve} the partial Q-value from the plane. The final feature-action Q-value is computed \rbc{as the \emph{sum of all}} the partial Q-values from \rbc{all} planes, \rbc{as shown in Fig.~\ref{fig:ke_design}(b)}.
\rbc{The use of tile coding provides two key advantages to Pythia. First, the tile coding of a feature enables the sharing of partial Q-values between similar feature values, which shortens prefetcher training time. Second, multiple planes reduces the chance of sharing partial Q-values between widely different feature values.}

\subsubsection{\textbf{Pipelined Organization of QVStore Search}}\label{sec:design_config_pipeline}
\rbc{To generate a} prefetch request, Pythia has to \rbc{(1)} look up the QVStore with the state-vector extracted from the current demand request, and (2) search for the action that has the maximum state-action Q-value (\circled{3} in Fig.~\ref{fig:overview}). As a result, the search operation lies on Pythia's critical path and directly impacts Pythia's prediction latency. To improve the prediction latency, we pipeline the search operation.

\begin{figure}[!h]
\vspace{-1em}
\centering
\includegraphics[scale=0.26]{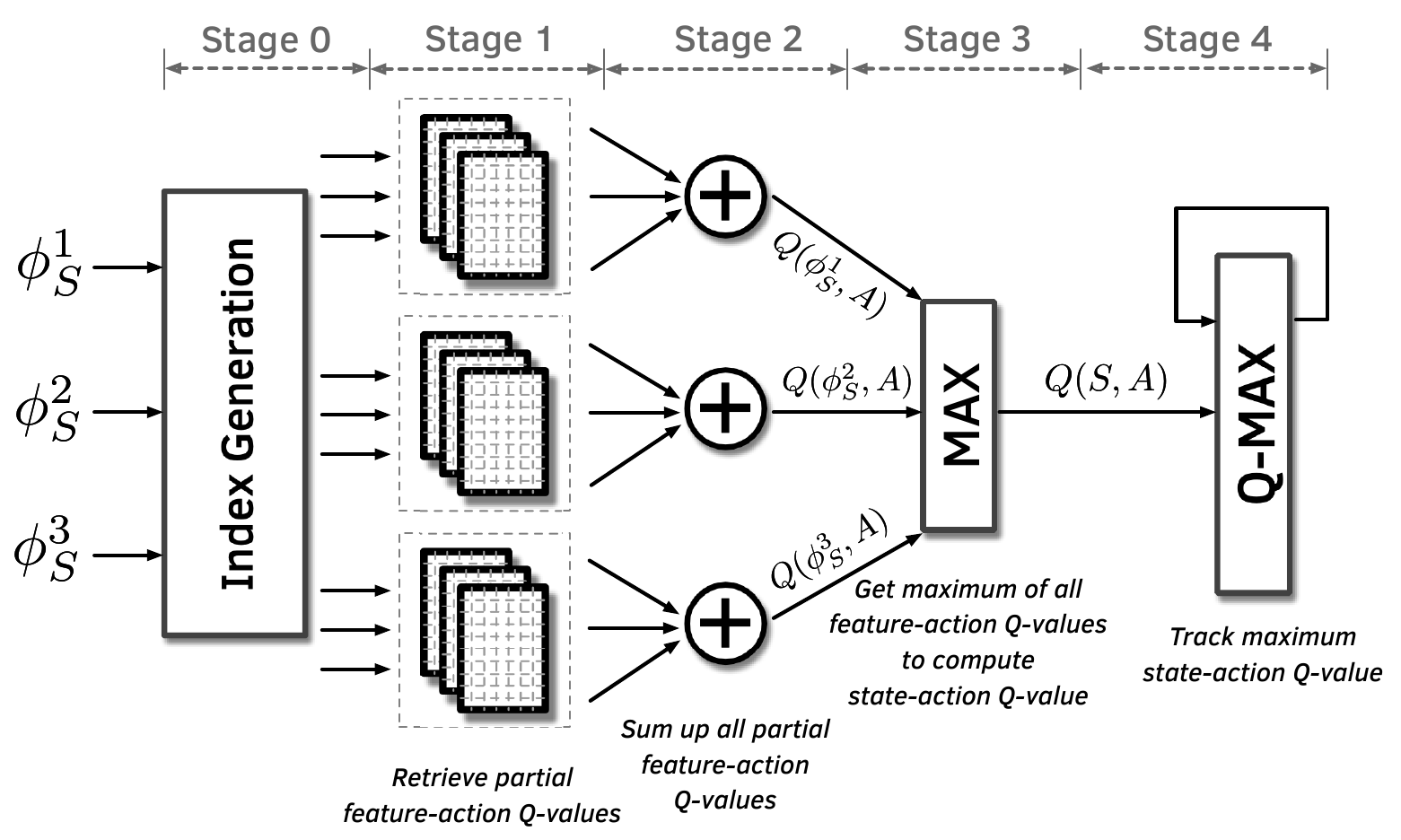}
\vspace{-1em}
\caption{\rbc{Pipelined} organization of QVStore search operation. The illustration depicts three program features, each having three planes. 
}
\label{fig:pythia_pipeline}
\vspace{-2em}
\end{figure}


\rbc{The Q-value search operation is implemented in the following way. For a given state-vector, Pythia iteratively retrieves the Q-value of each action. Pythia also maintains a variable, $Q_{max}$, that tracks the maximum Q-value found so far. $Q_{max}$ gets compared \rbc{to} every retrieved Q-value. The search operation concludes when Q-values for \emph{all} \rbc{possible} actions \rbc{have} been retrieved.}
We pipeline the search operation into five stages \rbc{as Fig.~\ref{fig:pythia_pipeline} shows}.
Pythia first computes the index for each plane and each constituent feature \rbc{of the given state-vector} \rbc{(Stage 0)}. \rbc{\rbc{In \rbc{Stage} 1}, Pythia uses the feature indices} and an action index to retrieve \rbc{the} partial Q-values from each plane. In \rbc{Stage 2}, Pythia \rbc{sums up} the partial Q-values to get the feature-action Q-value for each constituent feature. \rbc{In Stage 3}, Pythia computes \rbc{the maximum of all feature-action Q-values to get the state-action Q-value}. In \rbc{Stage} 4, \rbc{the maximum state-action Q-value found so far} is compared against the retrieved state-action Q-value, and the maximum Q-value is updated. Stage 2 (i.e., the partial Q-value summation) \rbc{is} the longest stage \rbc{of the pipeline and thus it} dictates the pipeline's throughput. 
\rbc{We accurately measure the area and power overhead of the pipelined implementation of the search operation by modeling Pythia using Chisel~\cite{chisel} hardware design language and synthesize the model using Synopsys design compiler~\cite{synopsys_dc} and 14-nm library from GlobalFoundries~\cite{global_foundries} (\cref{sec:eval_overhead}).}

\subsubsection{\textbf{Assigning Rewards and Updating Q-values}} \label{sec:ke_update}
\rbc{
To track usefulness of the prefetched requests, Pythia maintains a first-in-first-out list of recently taken actions, along with their corresponding prefetch addresses in EQ.
\emph{Every} prefetch action 
is inserted into EQ. A reward gets assigned to every EQ entry before or \rbc{when} it gets evicted from EQ.
During eviction, the reward and the state-action pair \rbc{associated with} the evicted EQ entry are used to update \rbc{the corresponding} Q-value in QVStore (\circled{6} in Fig.~\ref{fig:overview}). 
}

\rbc{We describe how Pythia appropriately assigns rewards to each EQ entry.
We divide the reward assignment into three classes based on \emph{when} the reward gets assigned to an entry: 
(1) immediate reward assignment during EQ insertion, (2) reward assignment during EQ residency, and (3) reward assignment during EQ eviction.
}
\rbc{If Pythia selects \rbc{the} action \rbc{\emph{not to prefetch}} or one that generates a prefetch request beyond the current physical page, Pythia immediately assigns \rbc{a} reward to the EQ entry. \rbc{For} out-of-page prefetch action, Pythia assigns $\mathcal{R}_{CL}$. \rbc{For the action \emph{not to prefetch}}, Pythia assigns $\mathcal{R}_{NP}^{H}$ or $\mathcal{R}_{NP}^{L}$, \rbc{based on whether} the current system memory bandwidth usage is high or low.
If \rbc{the} address of a demand request matches with the prefetch address stored in an EQ entry during its residency,
Pythia assigns $\mathcal{R}_{AT}$ or $\mathcal{R}_{AL}$ based on the \emph{filled} bit of the EQ entry. If the filled bit is set, it indicates that the demand request is generated \emph{after} the prefetch fill. Hence the prefetch is accurate and timely, and Pythia assigns the reward $\mathcal{R}_{AT}$. Otherwise, Pythia assigns the reward $\mathcal{R}_{AL}$.
If a reward does not get assigned to an EQ entry \rbc{until it is going to be evicted}, it signifies that the corresponding prefetch address is not yet demanded by the processor. Thus, Pythia assigns a reward $\mathcal{R}_{IN}^{H}$ or $\mathcal{R}_{IN}^{L}$ to the entry during eviction \rbc{based on whether} the current system memory bandwidth usage is high or low.  
}

\subsection{Automated Design-Space Exploration}\label{sec:tuning}

\rbc{We} propose an automated, performance-driven approach to systematically explore Pythia's vast design space and 
derive a \rbcb{basic} configuration\footnote{Using a compute-grid with ten $28$-core machines, the automated exploration across $150$ workload traces (mentioned in detail in \cref{sec:methodology}) takes $44$ hours to complete.} with appropriate program features, action set, reward and hyperparameters.
\rbc{Table~\ref{tab:pythia_config} shows the \rbcb{basic} configuration.}

\begin{table}[h]
    \vspace{-0.5em}
    \centering
    \footnotesize 
    \caption{\rbc{Basic} Pythia \rbc{configuration} derived from \rbc{our} automated design-space exploration}
    \vspace{-0.5em}
    \small
    \begin{tabular}{m{9.5em}m{18em}}
         \thickhline
         \textbf{Features} & \texttt{PC+Delta}, \texttt{Sequence of last-4 deltas}\\
         \hline
         \textbf{Prefetch Action List} & \{-6,-3,-1,0,1,3,4,5,10,11,12,16,22,23,30,32\} \\
         \hline
         \textbf{Reward Level Values} & $\mathcal{R}_{AT}$=$20$, $\mathcal{R}_{AL}$=$12$, $\mathcal{R}_{CL}$=$-12$, $\mathcal{R}_{IN}^{H}$=$-14$, $\mathcal{R}_{IN}^{L}$=$-8$, $\mathcal{R}_{NP}^{H}$=$-2$, $\mathcal{R}_{NP}^{L}$=$-4$\\
         \hline
         \textbf{Hyperparameters} & $\alpha=0.0065$, $\gamma=0.556$, $\epsilon=0.002$ \\
         \thickhline
    \end{tabular}
    \label{tab:pythia_config}
    \vspace{-1em}
\end{table}

\subsubsection{\textbf{Feature Selection}}\label{sec:tuning_feature_selection}
\rbc{We derive a list of possible program features for feature-space exploration in \rbc{four} steps. First, we derive a list of $4$ control-flow components, and $8$ data-flow components, which are mentioned in Table~\ref{table:list_of_features}. \rbc{Second,} we combine each control-flow component with each data-flow component \rbc{with the} concatenation operation, to \rbc{obtain} a total of $32$ possible program features.}
\rbc{Third,} we use \rbc{the} linear regression technique~\cite{lr1,lr2,lr3} to create any-one, any-two, and any-three feature-combinations from the set of $32$ initial features, each providing a different state-vector. \rbc{Fourth,} we run Pythia with every state-vectors across all single-core workloads (\cref{sec:methodology}) and select the \rbc{winning} state-vector that provides the highest performance gain over no-prefetching baseline. \rbc{As Table~\ref{tab:pythia_config} shows}, the two constituent features of the winning state-vector are \texttt{PC+Delta} and \texttt{Sequence of last-4 deltas}.

\begin{table}[htbp]
  \centering
  \footnotesize
  \caption{List of program control-flow and data-flow components used to derive the list of features for exploration}
  \vspace{-1em}
  \small
    \begin{tabular}{m{13.8em}m{13.8em}}
    \toprule
    \textbf{\footnotesize Control-flow Component} & \textbf{\footnotesize Data-flow Component} \\
    \midrule
    \begin{minipage}{13.7em}
      \footnotesize
      \vskip 14pt
      \begin{enumerate}[leftmargin=1.8em]
        \item PC of load request
        \item PC-path (XOR-ed last-3 PCs)
        \item PC XOR-ed branch-PC
        \item None
      \end{enumerate}
      \vskip 4pt
    \end{minipage} & 
    \begin{minipage}{13.7em}
      \footnotesize
      \vskip 1pt
      \begin{enumerate}[leftmargin=1.8em]
        \item Load cacheline address
        \item Page number
        \item Page offset
        \item Load address delta
        \item Sequence of last-4 offsets
        \item Sequence of last-4 deltas
        \item Offset XOR-ed with delta
        \item None
      \end{enumerate}
      \vskip 4pt
    \end{minipage}\\
    \bottomrule
    \end{tabular}%
  \label{table:list_of_features}%
  \vspace{-1em}
\end{table}%

\textbf{Rationale behind the \rbc{winning} state-vector.} The \rbc{winning} state-vector is intuitive as \rbc{its} constituent features \texttt{PC+Delta} and \texttt{Sequence of last-4 deltas} closely match with the program features exploited by \rbc{two prior state-of-the-art prefetchers}, Bingo~\cite{bingo} and SPP~\cite{spp}, respectively.
However, concurrently running SPP and Bingo \rbc{as a hybrid prefetcher does not provide} the same \rbc{performance} benefit as Pythia, as we show in \cref{sec:eval_perf_1T}. This is because combining SPP with Bingo not only \rbc{improves} their prefetch coverage, but also combines \rbc{their} prefetch overpredictions, \rbc{leading to performance degradation, especially in resource-constrained systems}. \rbc{In contrast}, Pythia's RL-based learning strategy \rbc{that inherently uses} the same two features successfully \rbc{increases} prefetch coverage, while maintaining \rbc{high prefetch accuracy}. As a result, \emph{Pythia not only outperforms SPP and Bingo individually, but also outperforms \rbc{the} combination of \rbc{the two} prefetchers}.

\subsubsection{\textbf{Action Selection}}\label{sec:tuning_action_selection}
In a system with \rbc{conventionally}-sized $4$KB pages and $64$B cachelines, Pythia's list of actions \rbc{(i.e., the list of possible prefetch offsets)} contains \emph{all} prefetch offsets in the range of $[-63,63]$. \rbc{However, such} a long action list poses two drawbacks. First, a long action list requires \rbc{more online exploration} to find the \rbc{best} prefetch offset \rbc{given a state-vector, \rbc{thereby} reducing Pythia's performance benefits}.
Second, a longer \rbc{action} list increases Pythia's storage \rbc{requirements}. 
\rbc{To avoid these problems, we prune the action list. We drop each action individually from the full action list $[-63,63]$ and measure the performance improvement relative to the performance improvement with the full action list, across all single-core workload traces. We prune any action that \emph{does not} have any significant impact on the performance.}
\rbc{Table~\ref{tab:pythia_config} shows} the final pruned action list.

\subsubsection{\textbf{Reward and Hyperparameter Tuning}}\label{sec:tuning_reward_hyp_selection}
\begin{sloppypar}
We separately tune \rbc{seven} reward \rbc{level values (i.e., $\mathcal{R}_{AT}$, $\mathcal{R}_{AL}$, $\mathcal{R}_{CL}$, $\mathcal{R}_{IN}^{H}$, $\mathcal{R}_{IN}^{L}$, $\mathcal{R}_{NP}^{H}$, and $\mathcal{R}_{NP}^{L}$) and three hyperparameters (i.e., learning rate $\alpha$, discount factor $\gamma$, and exploration rate $\epsilon$)} 
in three steps. 
\rbc{First,} we create a test trace suite by \emph{randomly} selecting $10$ \rbc{workload} traces from \rbc{all of our} $150$ \rbc{workload} traces (\cref{sec:methodology}). 
\rbc{Second}, \rbc{we create a list of tuning configurations} using \rbc{the} uniform grid search technique~\cite{grid_search1,grid_search2}. \rbc{To do so,} we first define a value range for each parameter to be tuned and divide the \rbc{value} range into uniform grids. For example, each of the three hyperparameters ($\alpha$, $\gamma$, and $\epsilon$) can take \rbc{a} value in the range of $[0,1]$. We divide each hyperparameter \rbc{range} into ten \rbc{exponentially}-sized grids (i.e., $1e^0$, $1e^{-1}$, $1e^{-2}$, etc.) to \rbc{obtain} $10\times10\times10=1000$ possible tuning configurations. 
\rbc{For each tuning configuration,} we run Pythia on the test trace suite and select the top-$25$ highest-performing configurations for the third step.
\rbc{Third}, we run Pythia on all single-core \rbc{workload} traces using each of the $25$ selected configurations. 
We select the \rbc{winning} configuration that provides the highest \rbc{average} performance gain.
\rbc{Table~\ref{tab:pythia_config} provides}
\rbc{reward level and hyperparameter values of \rbc{the \rbcb{basic} Pythia}.}
\end{sloppypar}

\subsection{Storage Overhead}
\rbc{Table~\ref{table:pythia_storage} shows the storage overhead of Pythia in its \rbcb{basic} configuration. Pythia requires only $25.5$KB of metadata storage. QVStore consumes $24$KB to store all Q-values. The EQ consumes only $1.5$KB.}

\begin{table}[h]
  \centering
  \footnotesize
  \caption{Storage overhead of Pythia}
  \vspace{-0.5em}
  \small
    \begin{tabular}{|m{4em}||m{18.7em}|c|}
    \hline
    \textbf{\footnotesize Structure} & \textbf{\footnotesize Description} & \textbf{\footnotesize Size} \\
    \hline
    \hline
    \begin{minipage}{4em}
    \footnotesize
    \vspace{11pt}
    \textbf{QVStore}
    \end{minipage}
    &
    \begin{minipage}{19em}
      \footnotesize
      \vskip 2pt
      \begin{itemize}[leftmargin=1em]
        \item \# vaults = $2$
        \item \# planes in each vault = $3$
        \item \# entries in each plane = feature dimension ($128$) $\times$ action dimension ($16$)
        \item Entry size = Q-value width ($16$b)
      \end{itemize}
      \vskip 2pt
    \end{minipage} &
    \begin{minipage}{3em}
    \footnotesize
    \vspace{16pt}
    \textbf{24 KB} 
    \end{minipage}
    \\
    \hline
    \begin{minipage}{3em}
    \footnotesize
    \vspace{8pt}
    \textbf{EQ}
    \end{minipage} &
    \begin{minipage}{19em}
      \footnotesize
      \vskip 2pt
      \begin{itemize}[leftmargin=1em]
        \item \# entries = 256
        \item Entry size = state ($21$b) + action index ($5$b) + reward ($5$b) + filled-bit ($1$b) + address ($16$b)
      \end{itemize}
      \vskip 2pt
    \end{minipage} &
    \begin{minipage}{3em}
    \footnotesize
    \vspace{10pt}
    \textbf{1.5 KB} 
    \end{minipage} \\    
    \hline
    \hline
    \textbf{\footnotesize Total} &      & \textbf{\footnotesize 25.5 KB} \\
    \hline
    \end{tabular}%
  \label{table:pythia_storage}%
  \vspace{-2em}
\end{table}

\subsection{Differences \rbc{from} Prior Work} \label{sec:compare_with_context}
The key idea of using RL in prefetching has been previously explored by \rbc{the} \rbc{\emph{context prefetcher} (CP)}~\cite{peled_rl}. Pythia significantly differs from it both in terms of (1) design \rbc{principles} (i.e., the reward, state, and action definition) and (2) the implementation.

\textbf{Reward.} \rbc{CP} naively defines the agent's reward as a continuous function of prefetch timeliness. Pythia not only considers coverage, accuracy, and timeliness but also system-level feedback like \rbc{memory} bandwidth usage to define discrete reward levels. This reward definition provides two key advantages to Pythia. First, unlike \rbc{CP}, Pythia can adaptively \rbc{trade off} prefetch coverage for accuracy (and \rbc{vice versa}) based on memory bandwidth usage. 
Second, one can easily customize \rbc{Pythia's} objective 
by changing the reward \rbc{values} via configuration registers to extract \rbc{even} higher performance. 

\textbf{State.} \rbc{CP} relies on \rbc{compiler-generated} hints \rbc{in its} state information. In contrast, Pythia extracts program features purely from hardware (e.g., PC, cacheline delta). 
\rbc{Thus, Pythia requires no changes to software and it is easier to adopt into existing microprocessors.} 

\textbf{Action.} Unlike Pythia, \rbc{CP} uses a full cacheline address as the agent's action. 
\rbc{The use of full cacheline address as action dramatically increases} the action space, \rbc{which results \rbc{higher storage cost}, longer training time, and reduced performance benefits.}

\begin{sloppypar}
\textbf{Implementation.}
Pythia's implementation differs \rbc{largely} from CP in two major ways. 
First, \rbc{CP} uses the contextual-bandit (CB) algorithm~\cite{chu2011contextual}, a simplified version of RL. 
The key difference between CB and RL is that a CB-solver cannot take \rbc{its} actions' \rbc{\emph{long-term}} consequences into account \rbc{when} selecting an action.
In contrast, RL-based Pythia weighs each probable prefetch action not only based on the expected immediate reward but also its long-term consequences (e.g., increased bandwidth usage or reduced prefetch accuracy in future)~\cite{rl_bible}. \rbc{As such,} Pythia provides more robust \rbc{and \rbc{far-sighted}} predictions than the \rbc{myopic} CB-based \rbc{CP}.
Second, Pythia organizes the Q-value storage \rbc{into} multiple vaults, each consisting of multiple planes. This hierarchical QVStore structure (1) enables pipelining the Q-value lookup to achieve high-throughput and low-latency prediction, and (2) easily scales out to support longer state-vectors by \rbc{simply} adding more vaults. 
\end{sloppypar}
\section{Methodology} \label{sec:methodology}

We use \rbc{the trace-driven} ChampSim simulator~\mbox{\cite{champsim}} to evaluate Pythia \rbc{and compare it to five} prior prefetching proposals.
We simulate an Intel Skylake~\mbox{\cite{skylake}}-like multi-core processor that supports \rbc{up to} 12 cores. \rbc{Table~\ref{table:sim_params} provides the \rbc{key system} parameters.}
For single-core simulations ($1C$), we warm up the core using $100$~M instructions \rbc{from each workload} and simulate the next $500$~M instructions. For multi-core multi-programmed simulations ($nC$), we use $50$~M and $150$~M instructions \rbc{from each workload} respectively to warmup and simulate. If a core finishes early, the workload is replayed \rbc{until} every core finishes simulating $150$~M instructions. 
We also \rbc{implement} Pythia using \rbc{the} Chisel~\cite{chisel} \rbc{hardware design language (HDL)} and functionally verify the resultant \rbc{register transfer logic (RTL) design} to accurately measure Pythia's \rbc{chip} area and power overhead.
The source-code of Pythia is freely available at~\cite{pythia_github}.

\begin{table}[h]
    \centering
    \footnotesize 
    \caption{\rbc{Simulated system} parameters}
    \vspace{-0.5em}
    \small
    \begin{tabular}{m{5.1em}m{22.7em}}
         \thickhline
         \textbf{\footnotesize Core} & {\footnotesize 1-12 cores, 4-wide OoO, 256-entry ROB, \rbc{72/56-entry LQ/SQ}}\\
         \hline
         \textbf{\footnotesize Branch \rbc{Pred.}} & {\footnotesize Perceptron-based~\cite{perceptron}, 20-cycle misprediction penalty} \\
         \hline
         \textbf{\footnotesize L1/L2 Caches} & {\footnotesize Private, 32KB/256KB, 64B line, 8 way, LRU, 16/32 MSHRs, 4-cycle/14-cycle round-trip latency} \\
         \hline
         \textbf{\footnotesize LLC} & {\footnotesize 2MB/core, 64B line, 16 way, SHiP~\cite{ship}, 64 MSHRs per LLC Bank, 34-cycle round-trip latency}\\
         \hline
         \begin{minipage}{7em}
         \footnotesize 
         \vspace{8pt}
         \textbf{Main Memory} 
         \end{minipage}
         & 
         \begin{minipage}{22.8em}
            \footnotesize 
            \vspace{2pt}
            \textbf{1C:} Single channel, 1 rank/channel; \textbf{4C:} Dual channel, 2 ranks/channel; \textbf{8C:} Quad channel, 2 ranks/channel; \\ 8 banks/rank, 2400 MTPS, 64b data-bus/channel, 2KB row buffer/bank, tRCD=15ns, tRP=15ns, tCAS=12.5ns
            \vspace{2pt}
         \end{minipage}\\
         \thickhline
    \end{tabular}
    \label{table:sim_params}
\end{table}

\subsection{Workloads}
\begin{sloppypar}
\rbc{We evaluate Pythia using a diverse set of memory-intensive workloads} spanning \texttt{SPEC CPU2006}~\cite{spec2006}, \texttt{SPEC CPU2017}~\cite{spec2017}, \texttt{PARSEC 2.1}~\cite{parsec}, \texttt{Ligra}~\cite{ligra}, and \texttt{Cloudsuite}~\cite{cloudsuite} \rbc{benchmark suites}.
\rbc{For \texttt{SPEC CPU2006} and \texttt{SPEC CPU20017} workloads, we reuse the instruction traces provided by \rbc{the} 2nd and \rbc{the} 3rd data prefetching championships (DPC)~\cite{dpc2, dpc3}. For \texttt{PARSEC} and \texttt{Ligra} workloads, we collect the instruction traces using \rbc{the} Intel Pin dynamic binary instrumentation tool~\cite{intel_pin}.
We do not consider workload traces that \rbc{have} lower than 3 last-level cache misses per kilo instructions (MPKI) in the baseline system with no prefetching. In all, we present results for $150$ workload traces spanning $50$ workloads. Table~\ref{table:workloads} shows a categorized view of all the workloads \rbc{evaluated} in this paper.}
For multi-core multi-programmed simulations, we create \rbc{both homogeneous and heterogeneous} trace mixes from our single-core trace list. For an $n$-core homogeneous trace mix, we run $n$ copies of \rbc{a} trace from our single-core trace list, one in each core. For \rbc{a} heterogeneous trace mix, we \emph{randomly} select $n$ traces from our single-core trace list and run one trace in every core. All the single-core traces and multi-programmed trace mixes used in our evaluation are freely available online~\cite{pythia_github}.
\end{sloppypar}

\begin{table}[htbp]
  \vspace{-0.5em}
  \centering
  \footnotesize 
  \caption{Workloads used for evaluation}
  \vspace{-0.8em}
  \small
    \begin{tabular}{m{3em}C{5em}C{2em}m{13em}}
    \toprule
    \textbf{\footnotesize Suite} & \multicolumn{1}{l}{\textbf{\footnotesize \# Workloads}} & \multicolumn{1}{l}{\textbf{\footnotesize \# Traces}} & \textbf{\footnotesize Example Workloads} \\
    \midrule
    {\footnotesize SPEC06} & {\footnotesize 16}    & {\footnotesize 28}    & {\footnotesize gcc, mcf, cactusADM, lbm, ...} \\
    {\footnotesize SPEC17} & {\footnotesize 12}    & {\footnotesize 18}    & {\footnotesize gcc, mcf, pop2, fotonik3d, ...} \\
    {\footnotesize PARSEC} & {\footnotesize 5}     & {\footnotesize 11}    & {\footnotesize canneal, facesim, raytrace, ...} \\
    {\footnotesize Ligra} & {\footnotesize 13}    & {\footnotesize 40}    & {\footnotesize BFS, PageRank, Bellman-ford, ...} \\
    {\footnotesize Cloudsuite} & {\footnotesize 4}     & {\footnotesize 53}    & {\footnotesize cassandra, cloud9, nutch, ...} \\
    \bottomrule
    \end{tabular}%
  \label{table:workloads}%
  \vspace{-1em}
\end{table}%

\subsection{Prefetchers}
\begin{sloppypar}
We \rbc{compare} Pythia \rbc{to five state-of-the-art} prior prefetchers: SPP~\cite{spp}, SPP+PPF~\cite{ppf}, SPP+DSPatch~\cite{dspatch}, Bingo~\cite{bingo}, and MLOP~\cite{mlop}.
We model each competing prefetcher using the source-code provided by their respective authors and fine-tune them in our environment to extract the \rbc{highest} performance gain \rbc{across} all single-core traces.
\rbc{Table~\mbox{\ref{table:pref_params}} shows the parameters of all evaluated prefetchers.}
Each prefetcher is trained on \rbc{L1-cache} misses and fills prefetched lines into L2 and LLC. 
\end{sloppypar}
We also \rbc{compare} Pythia against multi-level prefetchers found in commercial processors (e.g., stride prefetcher at L1-cache and streamer at L2~\cite{intel_prefetcher}) and IPCP~\cite{ipcp} in \cref{sec:eval_multi_level}. \rbc{For fair comparison, we add a simple PC-based stride prefetcher~\cite{stride,stride_vector,jouppi_prefetch} \rbc{at the} L1 \rbc{level}, along with Pythia at \rbc{the} L2 \rbc{level} for such \rbc{multi-level comparisons}.}

\begin{table}[h]
    \centering
    \footnotesize 
    \caption{Configuration of evaluated prefetchers}
    \vspace{-0.5em}
    \small
    \begin{tabular}{m{6em}|m{16.8em}||m{3em}}
    \thickhline
    {\footnotesize \textbf{SPP}~\cite{spp}} & {\footnotesize 256-entry ST, 512-entry 4-way PT, 8-entry GHR} & \textbf{\footnotesize 6.2~KB} \\
    \hline
    {\footnotesize \textbf{Bingo}~\cite{bingo}} & {\footnotesize 2KB region, 64/128/4K-entry FT/AT/PHT} & \textbf{\footnotesize 46~KB} \\
    \hline
    {\footnotesize \textbf{MLOP}~\cite{mlop}} & {\footnotesize 128-entry AMT, 500-update, 16-degree} & \textbf{\footnotesize 8~KB} \\
    \hline
    {\footnotesize \textbf{DSPatch}~\cite{dspatch}} & {\footnotesize Same configuration as in~\cite{dspatch}} & \textbf{\footnotesize 3.6~KB} \\
    \hline
    {\footnotesize \textbf{PPF}~\cite{ppf}} & {\footnotesize Same configuration as in~\cite{ppf}} & \textbf{\footnotesize 39.3~KB} \\
    \thickhline
    {\footnotesize \textbf{Pythia}} & {\footnotesize 2 features, 2 vaults, 3 planes, 16 actions} & \textbf{\footnotesize 25.5~KB}\\
    \thickhline
    \end{tabular}
    \label{table:pref_params}
    \vspace{-1em}
\end{table}

\section{Results}\label{sec:evaluation}

\subsection{Coverage and Overprediction in Single-core}\label{sec:eval_cov_acc}
\begin{figure}[!h]
\vspace{-1em}
\centering
\includegraphics[width=3.3in]{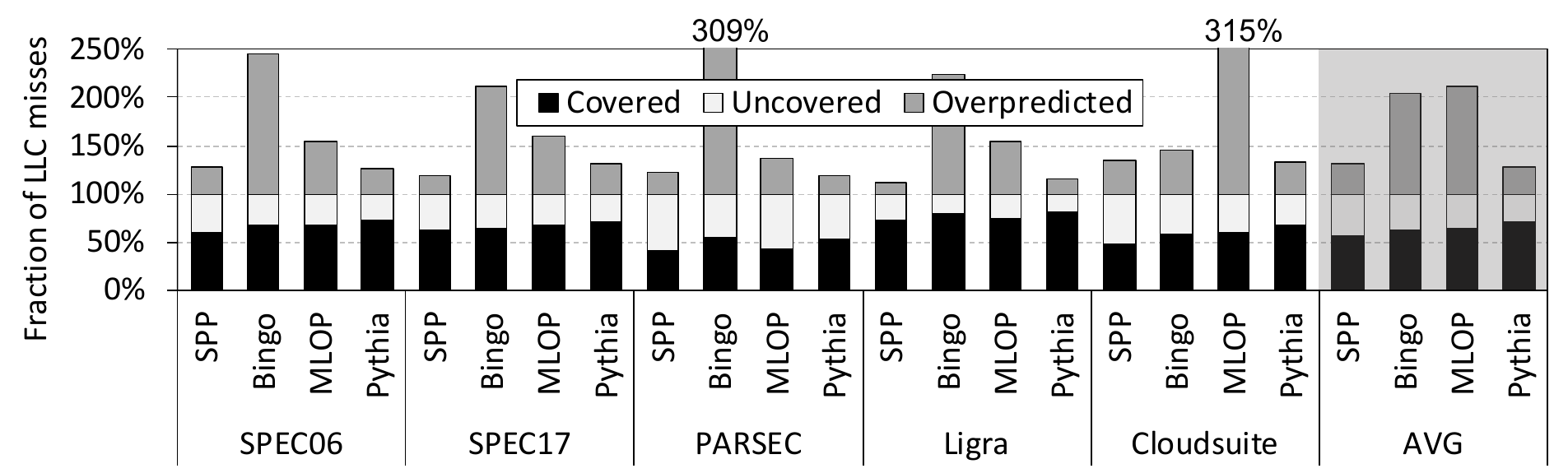}
\vspace{-0.5em}
\caption{Coverage and overprediction with respect to the baseline LLC misses in \rbc{the} single-core \rbc{system}. 
}
\label{fig:cov_acc}
\vspace{-1em}
\end{figure}

Fig.~\ref{fig:cov_acc} shows the coverage and overprediction of each prefetcher in \rbc{the} single-core \rbc{system}, \rbc{as} measured at the LLC-main memory boundary. 
The key takeaway is that Pythia improves prefetch coverage, while simultaneously reducing overprediction \rbc{compared to} state-of-the-art prefetchers. On average, Pythia provides $6.9$\%, $8.8$\%, and $14$\% \rbc{higher} coverage than MLOP, Bingo, and SPP respectively, while generating $83.8$\%, $78.2$\%, and $3.6$\% \rbc{fewer} overpredictions.   

\begin{figure*}[!h]
\centering
\includegraphics[scale=0.2257]{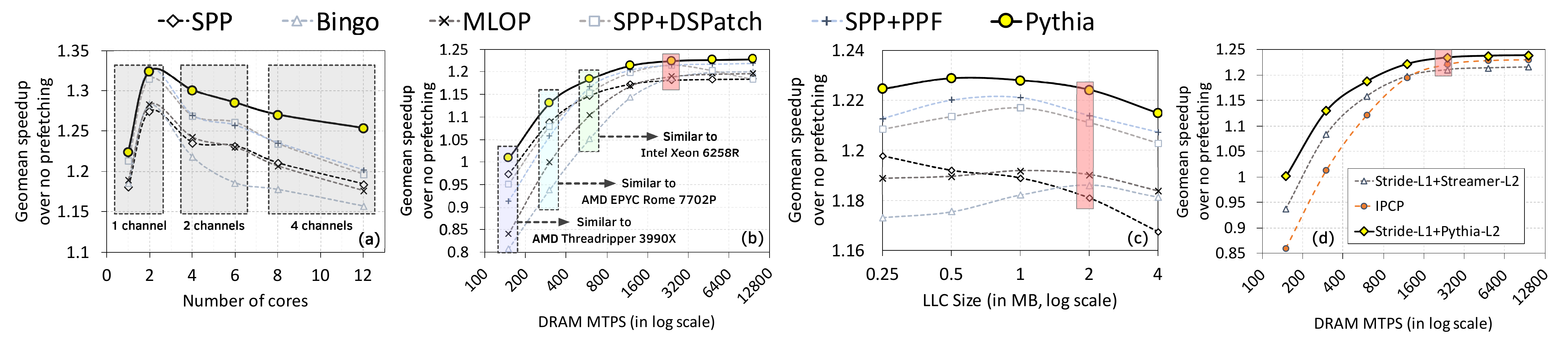}
\vspace{-2em}
\caption{Average performance improvement of prefetchers in \rbc{systems} with varying (a) number of cores, (b) DRAM \rbc{million transfers per second (MTPS)}, (c) LLC size, and (d) prefetching level. Each DRAM bandwidth configuration roughly matches MTPS/core of various commercial processors~\cite{intel_xeon_gold,zen_epyc,zen_threadripper}. The baseline bandwidth/LLC configuration is marked in red.}
\label{fig:scalability_master}
\vspace{-0.8em}
\end{figure*}

\subsection{Performance Overview}\label{sec:eval_scalability}
\subsubsection{\textbf{Varying Number of Cores}}

Figure~\mbox{\ref{fig:scalability_master}}(a) shows \rbc{the} performance improvement of all prefetchers averaged across all traces in single-core to 12-core \rbc{systems}. 
\rbc{To realistically model} modern commercial multi-core processors, we simulate $1$-$2$ core, $4$-$6$ core, and $8$-$12$ core \rbc{systems} with one, two, and four DDR4-$2400$ DRAM~\cite{ddr4} channels, respectively.
We make two key observations from Figure~\mbox{\ref{fig:scalability_master}}(a). First, Pythia consistently outperforms MLOP, Bingo, and SPP in \emph{all} \rbc{system} configurations. Second, Pythia's performance improvement over prior prefetchers increases \rbc{as} core count \rbc{increases}. 
In \rbc{the} single-core \rbc{system}, Pythia outperforms MLOP, Bingo, SPP, and an aggressive SPP with perceptron filtering (PPF~\mbox{\cite{ppf}}) by $3.4$\%, $3.8$\%, $4.3$\%, and $1.02$\% respectively.
In four (and twelve) core \rbc{systems}, Pythia outperforms MLOP, Bingo, SPP, and SPP+PPF by $5.8$\% ($7.7$\%), $8.2$\% ($9.6$\%), $6.5$\% ($6.9$\%), and $3.1$\% ($5.2$\%), respectively.


\subsubsection{\textbf{Varying DRAM Bandwidth}}
To evaluate Pythia in band-width-constrained, highly-\rbc{multi-}threaded commercial server-class processors, where each core \rbc{can have} only a fraction of a channel's bandwidth, we simulate the single-core single-channel configuration by scaling the DRAM bandwidth (Figure~\mbox{\ref{fig:scalability_master}}(b)). 
Each bandwidth configuration roughly \rbc{corresponds} to the available \rbc{per-core} DRAM bandwidth \rbc{in} various commercial processors \rbc{(e.g., Intel Xeon Gold~\cite{intel_xeon_gold}, AMD EPYC Rome~\cite{zen_epyc}, and AMD Threadripper~\cite{zen_threadripper})}.
The key takeaway is that Pythia \emph{consistently} outperforms all competing prefetchers in \emph{every} DRAM bandwidth configuration from $\frac{1}{16}\times$ to $4\times$ bandwidth of the baseline system.
Due to \rbc{their large overprediction \rbc{rates}}, \rbc{the} performance \rbc{gains} of MLOP and Bingo \rbc{reduce} sharply \rbc{as} DRAM bandwidth \rbc{decreases}. \rbc{By} actively trading off prefetch coverage for higher accuracy based on \rbc{memory} bandwidth usage, Pythia outperforms MLOP, Bingo, SPP, and SPP+PPF by $16.9$\%, $20.2$\%, $3.7$\%, and $9.5$\% respectively in the most bandwidth-constrained configuration with 150 million transfers per second (MTPS).
In \rbc{the} $9600$-MTPS configuration, every prefetcher enjoys ample DRAM bandwidth. Pythia \rbc{still} outperforms MLOP, Bingo, SPP, and SPP+PPF by $3$\%, $2.7$\%, $4.4$\%, and $0.8$\%, respectively.

\subsubsection{\textbf{Varying LLC Size}}
\begin{sloppypar}
Fig.~\ref{fig:scalability_master}(c) shows performance of all prefetchers averaged across all traces in \rbc{the} single-core \rbc{system} \rbc{while} varying \rbc{the} LLC size from $\frac{1}{8}\times$ to $2\times$ of the baseline 2MB LLC.
\rbc{The key takeaway is that Pythia consistently outperforms all prefetchers in \emph{every} LLC size configuration.}
For $256$KB (and $4$MB) LLC, Pythia outperforms MLOP, Bingo, SPP, and SPP+PPF by $3.6$\% ($3.1$\%), $5.1$\% ($3.4$\%), $2.7$\% ($4.8$\%), and $1.2$\% ($0.8$\%), respectively.
\end{sloppypar}
\subsubsection{\textbf{Comparison to Multi-level Prefetching Schemes}}\label{sec:eval_multi_level}

Figure~\mbox{\ref{fig:scalability_master}}(d) shows \rbc{the performance comparison of Pythia in single-core system with varying DRAM bandwidth against two state-of-the-art \emph{multi-level} prefetching schemes}: (1) stride prefetcher~\cite{stride,stride_vector,jouppi_prefetch} at L1 and streamer~\cite{streamer} at L2 cache found in commercial Intel processors~\cite{intel_prefetcher}, and (2) IPCP, the winner of \rbc{the} third data prefetching championship~\cite{dpc3}. For \rbc{fair} comparison, we add a stride prefetcher in \rbc{the} L1 cache along with Pythia in \rbc{the} L2 cache for this experiment and measure performance over the no prefetching baseline.
\rbc{The key takeaway is that \rbc{Stride+Pythia} consistently outperforms stride+streamer and IPCP in \emph{every} DRAM bandwidth configuration. Stride+Pythia outperforms \rbc{Stride+Streamer} and IPCP by $6.5$\% and $14.2$\% in \rbc{the} $150$-MTPS configuration and \rbc{by} $2.3$\% and $1.0$\% in \rbc{the} $9600$-MTPS configuration, respectively.}

\subsection{Performance Analysis}\label{sec:eval_perf}

\subsubsection{\textbf{Single-core}}\label{sec:eval_perf_1T}
\begin{figure}[!h]
\centering
\includegraphics[width=3.3in]{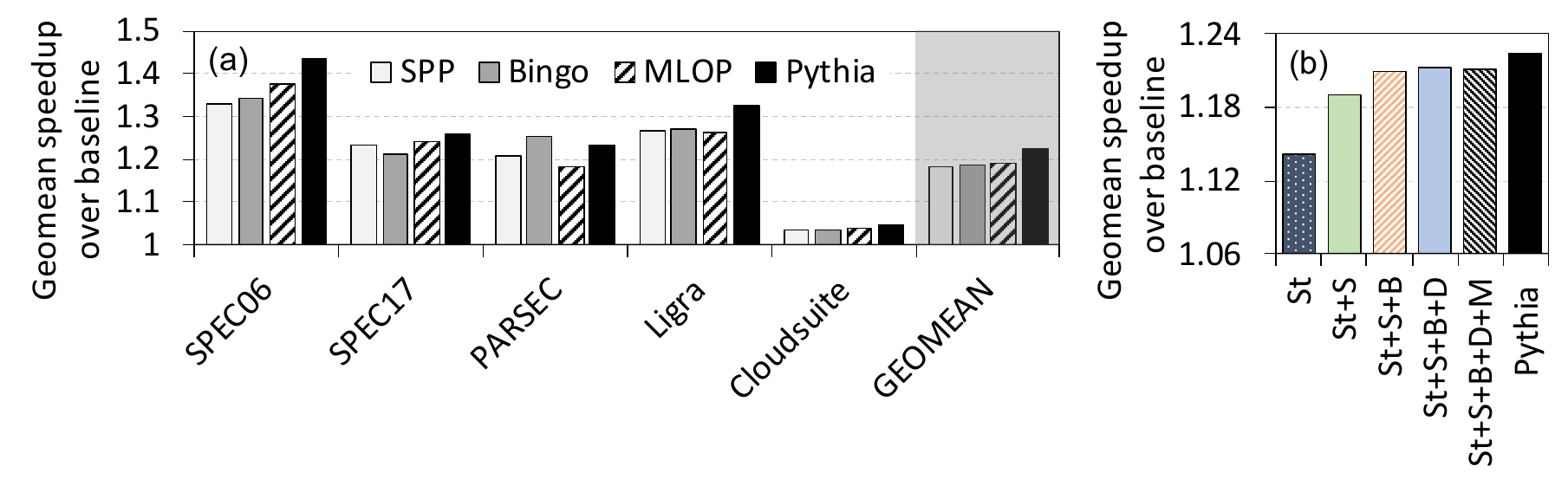}
\vspace{-1.2em}
\caption{Performance improvement in single-core workloads. St=Stride, S=SPP, B=Bingo, D=DSPatch, and M=MLOP.}
\label{fig:perf_1T}
\vspace{-1em}
\end{figure}

Fig.~\ref{fig:perf_1T}(a) \rbc{shows} the performance improvement of each individual prefetcher in each workload category in the single-core \rbc{system}. We make two major observations.
First, Pythia improves performance by $22.4$\% on average over a no-prefetching baseline. Pythia outperforms MLOP, Bingo, and SPP by $3.4$\%, $3.8$\%, and $4.3$\% on average, respectively. 
Second, \rbc{only} Bingo outperforms Pythia \rbc{only} in \rbc{the} \texttt{PARSEC} suite, by $2.3$\%. However, Bingo's performance comes at \rbc{the} cost of \rbc{a} high overprediction \rbc{rate}, which hurts performance in multi-core \rbc{systems (see \cref{sec:eval_perf_4T})}.

To demonstrate \rbc{the novelty of Pythia's RL-based prefetching approach using multiple program features}, Fig.~\ref{fig:perf_1T}(b) compares Pythia's performance improvement with the performance improvement of \rbc{various} combinations of prior prefetchers.
Pythia not only outperforms \rbc{all prefetchers (stride, SPP, Bingo, DSPatch, and MLOP)} individually, but also outperforms \rbc{their combination} by $1.4$\% on average, with less than half of the combined storage size \rbc{of the five prefetchers}. 
\rbc{We conclude that Pythia's RL-based prefetching approach using multiple program features under one single framework provides \rbc{higher} performance \rbc{benefit} than combining multiple prefetchers, each exploiting only one program feature.}

\subsubsection{\textbf{Four-core}}\label{sec:eval_perf_4T}
\begin{figure}[!h]
\centering
\includegraphics[width=3.3in]{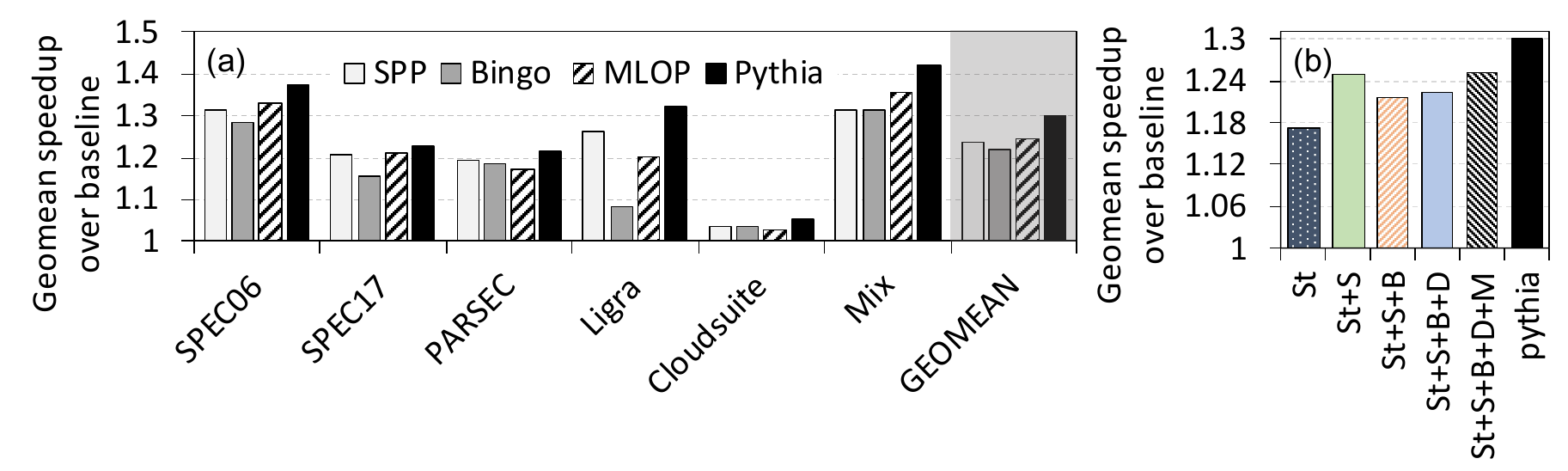}
\vspace{-2em}
\caption{Performance in \rbc{the} four-core \rbc{system}. 
}
\label{fig:perf_4T}
\vspace{-2em}
\end{figure}

Fig.~\ref{fig:perf_4T}(a) \rbc{shows} the performance improvement of each individual prefetcher in each workload category in the four-core \rbc{system}. We make \rbc{two} major observations. 
First, Pythia provides significant performance improvement over all prefetchers in \emph{every} workload category in \rbc{the} four-core \rbc{system}. On average, Pythia outperforms MLOP, Bingo, and SPP by $5.8$\%, $8.2$\%, and $6.5$\% respectively.
Second, unlike \rbc{in the} single-core \rbc{system}, Pythia outperforms Bingo in \texttt{PARSEC} by $3.0$\% in \rbc{the} four-core \rbc{system}. \rbc{This is due to Pythia's ability to dynamically increase prefetch accuracy during high DRAM bandwidth usage.}

\rbc{Fig.~\ref{fig:perf_4T}(b) shows that} Pythia outperforms the combination of stride, SPP, Bingo, DSPatch, and MLOP prefetchers by $4.9$\% on average.
Unlike \rbc{in the} single-core system, combining \rbc{more} prefetchers on top of stride+SPP in four-core system lowers the overall performance gain. 
This is due to the additive increase in the overpredictions \rbc{made} by each individual prefetcher, \rbc{which leads to performance degradation in the bandwidth-constrained four-core system.} Pythia's RL-based framework holistically learns to prefetch using multiple program features \rbc{and} \rbc{generates} \rbc{fewer} overpredictions, outperforming \rbc{all} combinations of all individual prefetchers.


\subsubsection{\textbf{Benefit of \rbc{Memory} Bandwidth \rbc{Usage} Awareness}} \label{sec:eval_bw_awareness}
\begin{figure}[!h]
\vspace{-1em}
\centering
\includegraphics[width=\columnwidth]{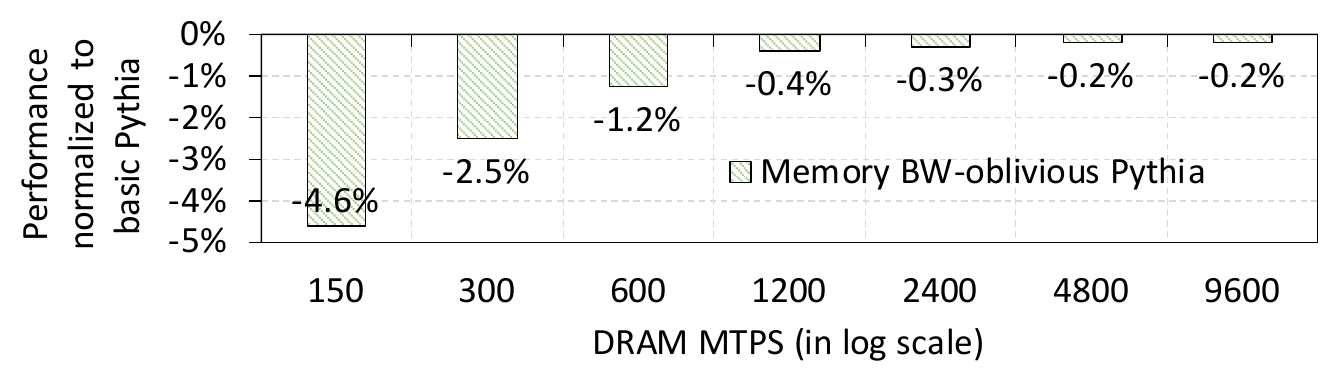}
\vspace{-2em}
\caption{Performance of \rbc{memory} bandwidth-oblivious Pythia \rbc{versus} the \rbcb{basic} Pythia.}
\label{fig:bw_awareness}
\vspace{-2em}
\end{figure}

To demonstrate the benefit of Pythia's awareness \rbc{of} system memory bandwidth usage, we compare the performance of the full-blown Pythia with a new version of Pythia that is oblivious to system memory bandwidth usage. We create this bandwidth-oblivious version of Pythia by \rbc{setting the high and low bandwidth usage variants of the rewards $\mathcal{R}_{IN}$ and $\mathcal{R}_{NP}$ to the same value (\rbc{i.e.,} essentially removing the bandwidth usage distinction from the reward values). More specifically, we set $\mathcal{R}_{IN}^{H}=\mathcal{R}_{IN}^{L}=-8$ and $\mathcal{R}_{NP}^{H}=\mathcal{R}_{NP}^{L}=-4$.}
Fig.~\ref{fig:bw_awareness} shows the performance benefit of the \rbc{memory} bandwidth-oblivious Pythia normalized to the \rbcb{basic} Pythia \rbc{as we vary the DRAM bandwidth}. \rbc{The key takeaway is that} 
\rbc{the bandwidth-oblivious Pythia loses performance by up to $4.6$\% on average across all \rbc{single-core} traces when the available memory bandwidth is low ($150$-MTPS to $600$-MTPS configuration).}
However, \rbc{when the available memory bandwidth is high ($1200$-MTPS to $9600$-MTPS)}, the \rbc{memory} bandwidth-oblivious Pythia \rbc{provides} similar performance improvement \rbc{to} the \rbcb{basic} Pythia.
We conclude that, memory bandwidth awareness gives Pythia the ability to provide robust performance \rbc{benefits} across a wide range of system \rbc{configurations}.


\subsection{Performance \rbc{on Unseen} Traces} \label{sec:eval_unknown_traces}
To demonstrate Pythia's ability to \rbc{provide performance gains} across \rbc{workload traces that are not used \rbc{at all} to tune Pythia}, we evaluate Pythia using an additional $500$ traces from the second value prediction championship~\mbox{\cite{cvp2}} \rbc{on} both single-core and four-core \rbc{systems}. These traces \rbc{are classified into} floating-point, integer, crypto, and server \rbc{categories} and each of them has at least 3 LLC MPKI in \rbc{the} baseline \rbc{without prefetching}. 
No prefetcher, including Pythia, has been tuned on these traces. 
\rbc{In \rbc{the} single-core system, Pythia outperforms MLOP, Bingo, and SPP on average by $8.3$\%, $3.5$\%, and $4.9$\%, respectively, across these traces. In \rbc{the} four-core system, Pythia outperforms MLOP, Bingo, and SPP on average by $9.7$\%, $5.4$\%, and $6.7$\%, respectively.}
\rbc{We conclude that, Pythia, tuned on a set of workload traces, provides \rbc{equally high} (or even better) performance benefits \rbc{on} unseen traces \rbc{for} which it has not \rbc{been} tuned.}

\begin{figure}[!h]
\vspace{-1em}
\centering
\includegraphics[scale=0.26]{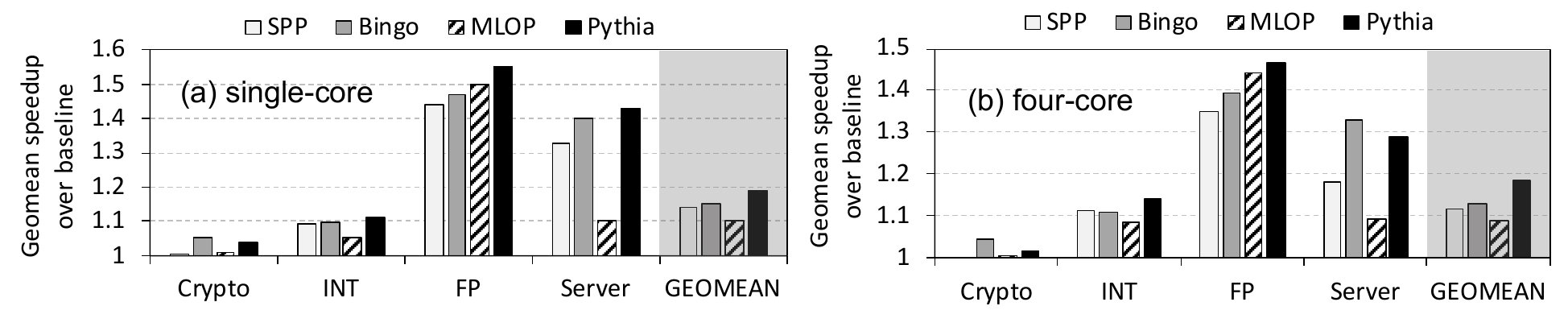}
\vspace{-2em}
\caption{Performance on unseen traces.
}
\label{fig:perf_cvp1}
\vspace{-2em}
\end{figure}


\subsection{Understanding Pythia Using a Case Study}\label{sec:eval_case_study}
\begin{figure}[!h]
\vspace{-1em}
\centering
\includegraphics[scale=0.22]{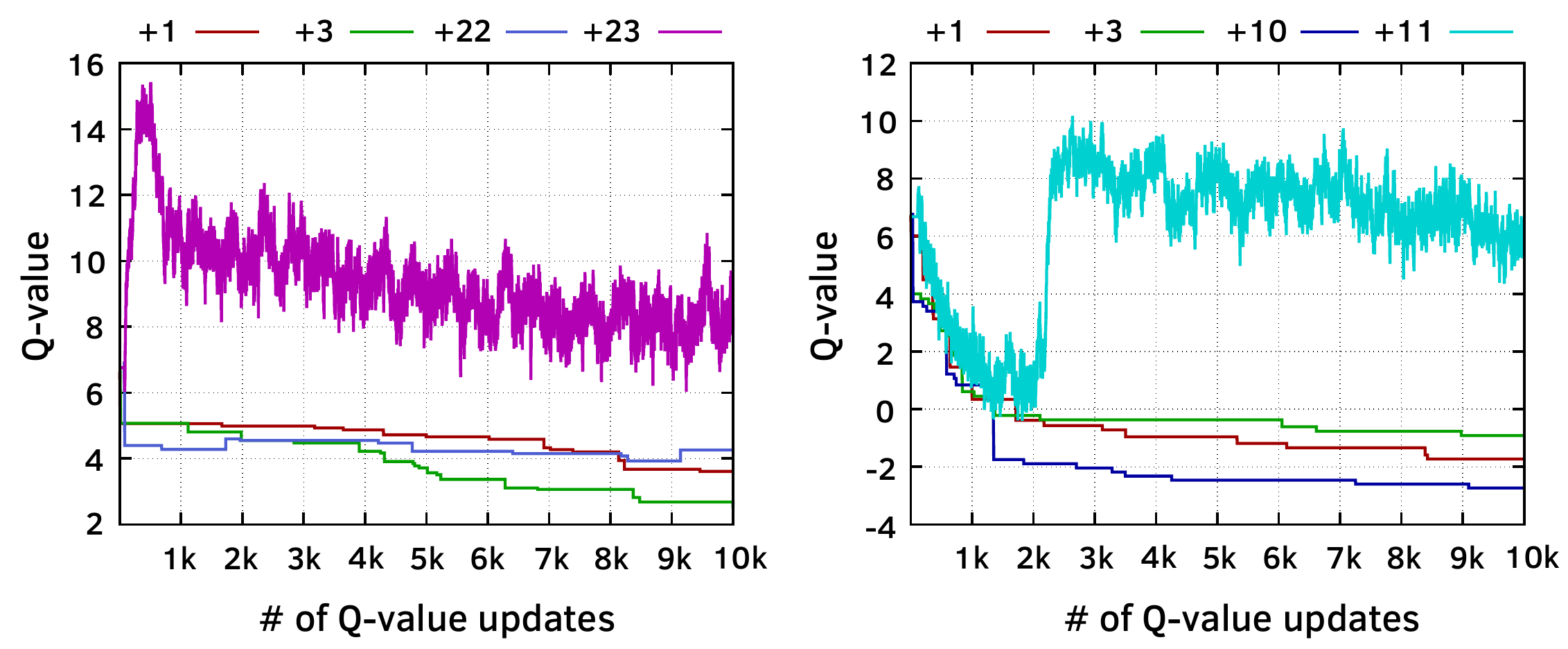}
\vspace{-1.8em}
\caption{Q-value curves of \texttt{PC+Delta} feature \rbc{values} (a) \texttt{0x436a81+0} and (b) \texttt{0x4377c5+0} in \texttt{459.GemsFDTD-1320B}.}
\label{fig:deepdive_gems}
\vspace{-1.5em}
\end{figure}

\begin{sloppypar}
\rbc{We} delve deeper into \rbc{an example} workload trace, \texttt{459.GemsFDTD-1320B}, from \texttt{SPEC CPU2006} suite \rbc{to provide more insight into Pythia's prefetching strategy and benefits}. 
In this trace, the top two most selected \rbc{prefetch} offsets by Pythia are $+23$ and $+11$, which cumulatively account for nearly $72$\% of all offset selections. 
For each of these offsets, we examine the program feature \rbc{value} that \rbc{selects} that offset \rbc{the most}. For simplicity, we only focus on the \texttt{PC+Delta} feature here.
The \texttt{PC+Delta} feature \rbc{values} \texttt{0x436a81+0} and \texttt{0x4377c5+0} select the \rbc{offsets} $+23$ and $+11$ the most, respectively. \rbc{Fig.~\ref{fig:deepdive_gems}(a) and (b)} show the Q-value curve of different actions for these feature. The x-axis shows the number of Q-value updates to the corresponding feature. Each color-coded line represents the Q-value of the respective action.
\end{sloppypar}

\begin{sloppypar}
As Fig.~\ref{fig:deepdive_gems}(a) shows, \rbc{the} Q-value of action $+23$ for feature \rbc{value} \texttt{0x436a81+0} consistently stays higher than \rbc{all} other actions (only three other \rbc{representative} actions are shown in~\ref{fig:deepdive_gems}(a)). This means Pythia actively \rbc{favors} to prefetch \rbc{using} $+23$ offset whenever the PC \texttt{0x436a81} generates the first load to a physical page \rbc{(hence the delta $0$)}. By dumping the program trace, we indeed find that whenever PC \texttt{0x436a81} generates the first load to a physical page, there is only one more \rbc{address demanded} in that page \rbc{that} is $23$ cachelines ahead from the first loaded cacheline. \rbc{In} this case, the positive reward for generating \rbc{a} correct prefetch with offset $+23$ drives the Q-value of $+23$ \rbc{much} higher than \rbc{those of} other offsets and Pythia \rbc{successfully uses} the offset $+23$ for \rbc{prefetch request generation given} the feature \rbc{value} \texttt{0x436a81+0}.
We see similar \rbc{a} trend for the feature \rbc{value} \texttt{0x4377c5+0} with offset $+11$ (Fig.~\ref{fig:deepdive_gems}(b)).
\end{sloppypar}

\subsection{Performance Benefits via Customization}\label{sec:eval_pythia_custom}
\rbc{In this section, we show two examples of Pythia's \rbc{online} customization ability to extract \rbc{even} higher performance gain than the baseline Pythia configuration in target workload suites. First, we customize Pythia's reward level values for \rbc{the} \texttt{Ligra} \rbc{graph processing} workloads. Second, we customize the program features used by Pythia for \rbc{the} \texttt{SPEC CPU2006} workloads.}

\subsubsection{\textbf{Customizing Reward Levels}}\label{sec:eval_pythia_custom_reward}
\begin{figure}[!h]
\vspace{-1em}
\centering
\includegraphics[width=3.3in]{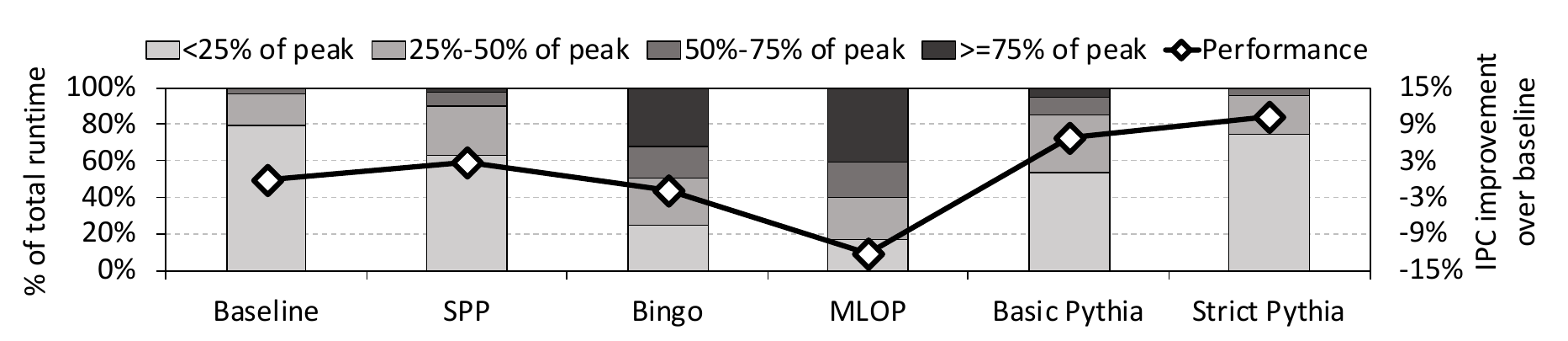}
\vspace{-1em}
\caption{Performance and main memory bandwidth usage of prefetchers in \texttt{Ligra-CC}.}
\label{fig:deepdive_ligra_cc}
\vspace{-1em}
\end{figure}

\begin{figure}[!h]
\vspace{-1em}
\centering
\includegraphics[width=3.3in]{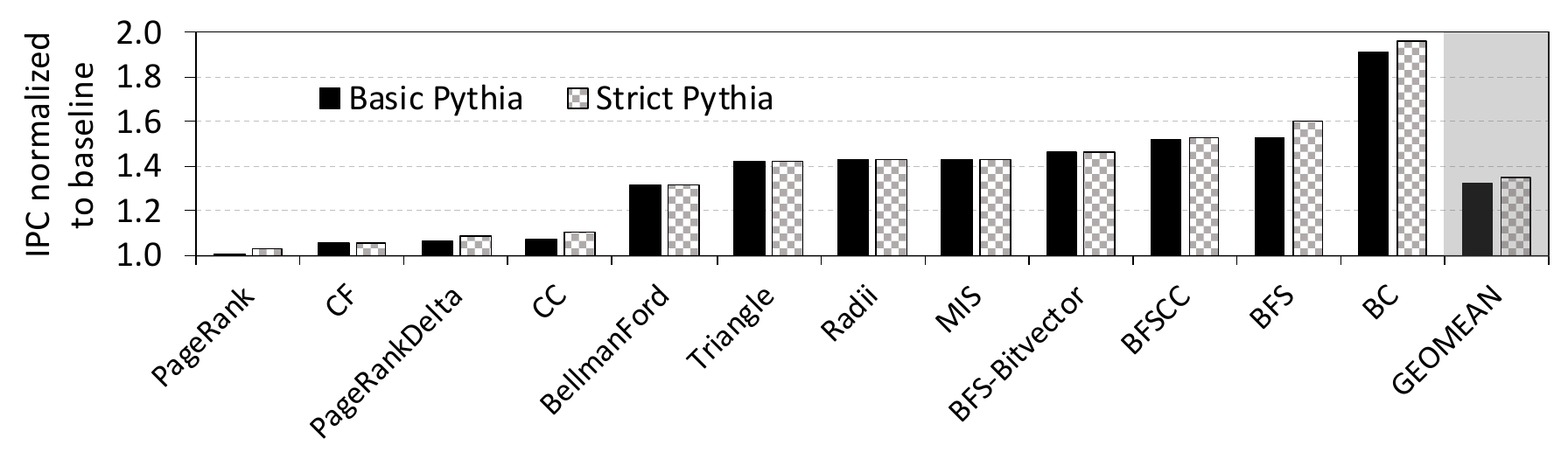}
\vspace{-1em}
\caption{Performance of the \rbc{basic} and strict \rbc{Pythia} configurations \rbcb{on} the \texttt{Ligra} workload suite.}
\label{fig:ligra}
\vspace{-1em}
\end{figure}

\begin{sloppypar}
For workloads from \rbc{the} Ligra \rbc{graph processing} suite, we observe a general trend that a prefetcher with higher prefetch accuracy typically provides higher performance benefits. This is because any incorrect prefetch request wastes precious main memory bandwidth, which is already heavily used by the \rbc{demand requests of the} workload.
Thus, to improve Pythia's performance benefit in \rbc{the} Ligra suite, we create a new \emph{strict configuration} of Pythia that favors \rbc{\emph{not to prefetch}} \rbc{over} \rbc{generating inaccurate prefetches}. We create this strict configuration by simply \rbc{reducing} the reward level values for inaccurate prefetch (i.e., $\mathcal{R}_{IN}^{H}=-22$ and $\mathcal{R}_{IN}^{L}=-20$) and \rbc{increasing the reward level values for} no prefetch (i.e., $\mathcal{R}_{NP}^{H}=\mathcal{R}_{NP}^{L}=0$). 
\end{sloppypar}
\rbc{Fig.~\ref{fig:deepdive_ligra_cc} shows the percentage of the total runtime the workload spends in different bandwidth usage buckets in primary y-axis and the overall performance improvement in the secondary y-axis for each competing prefetcher \rbc{in one example workload} from \rbc{the} \texttt{Ligra} suite, \texttt{Ligra-CC}.}
\rbc{We make two key observations.}
\rbc{First, with MLOP and Bingo prefetchers enabled, \texttt{Ligra-CC} spends \rbc{a much} higher percentage of runtime consuming more than half of the peak DRAM bandwidth than in the no prefetching baseline. As a result, MLOP and Bingo underperforms the no prefetching baseline by $11.8\%$ and $1.8\%$, respectively. In contrast, \rbcb{basic} Pythia \rbc{leads to only} a modest memory bandwidth usage overhead, and outperforms the no prefetching baseline by $6.9$\%.}
\rbc{Second,} in \rbc{the} strict configuration, Pythia \rbc{has even less} \rbc{memory} bandwidth \rbc{usage} overhead, and provides $3.5$\% \rbc{higher} performance \rbc{than} the \rbcb{basic} Pythia configuration ($10.4$\% over \rbc{the no prefetching} baseline), without any hardware changes.
Fig.~\ref{fig:ligra} shows the performance \rbc{benefits} of the \rbcb{basic} and strict Pythia configurations for all \rbc{workloads from} \texttt{Ligra}. The key takeaway is that by simply changing the reward level values via configuration registers on the silicon, \rbc{strict} Pythia \rbc{provides} up to $7.8$\% ($2.0$\% on average) \rbc{higher} performance \rbc{than} \rbcb{basic} Pythia.
\rbc{We conclude that the objectives of Pythia can be easily customized via simple configuration registers for target workload suites to extract \rbc{even} higher performance benefits, without any changes to the underlying hardware.}

\subsubsection{\textbf{\rbc{Customizing Feature Selection}}}\label{sec:eval_pythia_custom_feature}
\begin{figure}[!h]
\centering
\includegraphics[width=3.3in]{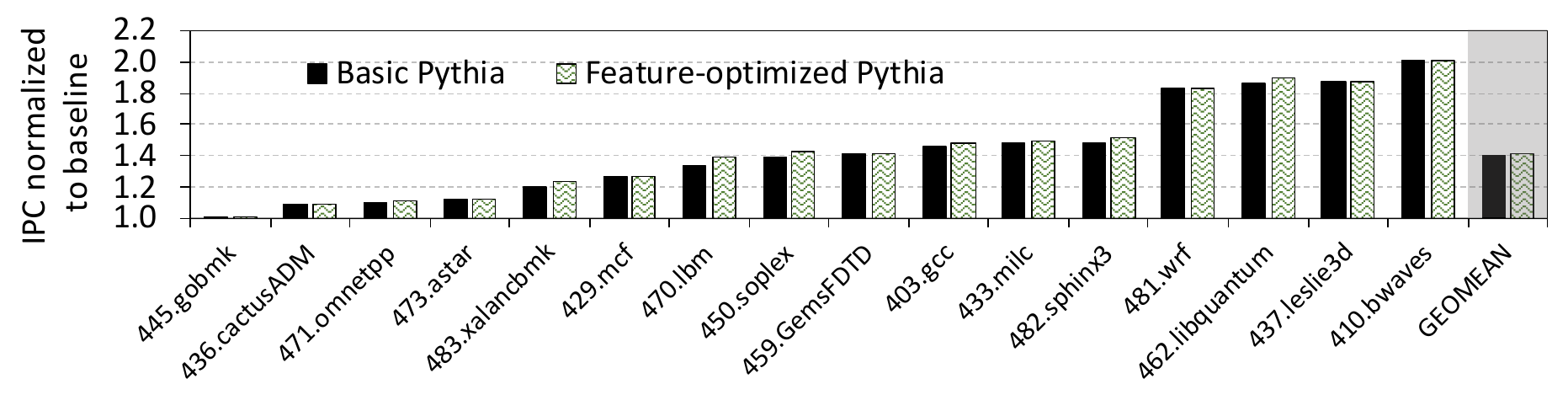}
\vspace{-1em}
\caption{Performance \rbc{of} \rbc{the basic} and \rbc{feature-optimized} Pythia \rbcb{on} \rbc{the} \texttt{SPEC CPU2006} suite.}
\label{fig:spec}
\vspace{-1.2em}
\end{figure}

To maximize the performance \rbc{benefits} of Pythia on \rbc{the} \texttt{SPEC CPU2006} workload suite, we run all one-combination and two-combination of program features from the initial set of $32$ supported features. For each workload, we fine-tune Pythia using the feature combination that provides the highest performance \rbc{benefit}. We call this the \emph{\rbc{feature-}optimized configuration} of Pythia for \texttt{SPEC CPU2006} suite. Fig.~\ref{fig:spec} shows the performance \rbc{benefits} of \rbcb{the basic} and optimized configurations of Pythia for all \texttt{SPEC CPU2006} workloads. The key takeaway is \rbc{that} by simply fine-tuning the program feature selection, Pythia delivers up to $5.1$\% ($1.5$\% on average) performance improvement on top of the \rbcb{basic} Pythia configuration.

\subsection{Overhead Analysis}\label{sec:eval_overhead}
\begin{sloppypar}
\rbc{To accurately estimate Pythia's chip area and power \rbc{overheads},} we implement \rbc{the full-blown} Pythia, including all fixed-point adders, multipliers, and the pipelined QVStore search operation (\cref{sec:design_config_pipeline}), using \rbc{the} Chisel~\cite{chisel} \rbc{hardware design language (HDL). \rbc{We} extensively verify the functional correctness of the resultant register transfer logic (RTL) design \rbc{and} synthesize the RTL design using Synopsys Design Compiler~\cite{synopsys_dc} and 14-nm library from GlobalFoundries~\cite{global_foundries} to estimate Pythia's area and power overhead.}
Pythia \rbc{consumes} $0.33$ mm2 of area and $55.11$ mW of power in each core. 
\rbc{The QVStore component}
consumes $90.4$\% and $95.6$\% of the \rbc{total} area and power \rbc{of Pythia,} respectively. 
\rbc{With respect to the overall die area and power consumption of a}
$4$-core desktop-class Skylake processor with \rbc{the} \rbc{lowest} TDP budget~\cite{skylake-4c}, \rbc{and a $28$-core server-class Skylake processor with the highest TDP budget}, \rbc{Pythia (implemented in all cores) incurs \rbc{area \& power} overheads of only $1.03$\% \& $0.4$\%, \rbc{and $1.33$\% \& $0.75$\%,} respectively. 
We conclude that Pythia's performance benefits come at a \rbc{very modest} cost in area and power \rbc{overheads} across a variety of commercial processors.
}
\end{sloppypar}


\begin{table}[htbp]
  \vspace{-0.8em}
  \centering
  \footnotesize
  \caption{Area and power \rbc{overhead} of Pythia}
  \vspace{-1.2em}
    \begin{tabular}{p{17em}C{4.2em}C{4.2em}}
    \multicolumn{3}{c}{\textbf{Pythia's area}: $0.33$ mm2/core; \textbf{Pythia's power}: $55.11$ mW/core\vspace{0.5em}} \\
    \toprule
    \textbf{Overhead \rbc{compared to real systems}} & \textbf{Area} & \textbf{Power} \\
    \midrule
    4-core Skylake D-2123IT, 60W TDP~\cite{skylake-4c} & 1.03\% & 0.37\% \\
    18-core Skylake 6150, 165W TDP~\cite{skylake-18c} & 1.24\% & 0.60\% \\
    28-core Skylake 8180M, 205W TDP~\cite{skylake-28c} & 1.33\% & 0.75\% \\
    \bottomrule
    \end{tabular}%
  \label{tab:overhead_analysis}%
  \vspace{-1.4em}
\end{table}%

\section{Other Related Works} \label{related_work}
\rbc{To our knowledge, Pythia is the first RL-based customizable prefetching framework that can learn to prefetch using multiple different program features and system-level feedback information inherent to its design, to provide performance benefits across a wide range of workloads and changing system configurations.}
\rbc{We already compare Pythia against five state-of-the-art prefetching proposals~\cite{spp,bingo,dspatch,ppf,mlop} in \cref{sec:evaluation}.}
In this section, we qualitatively compare \rbc{Pythia against other prior prefetching techniques.}


\begin{sloppypar}
\textbf{Traditional Prefetchers.} 
We divide the traditional prefetching algorithms \rbc{into} three broad categories: \rbc{precomputation}-based, temporal, and spatial. \rbc{Precomputation}-based prefetchers (e.g., runahead~\cite{dundas,mutlu2003runahead,mutlu2003runahead2, mutlu2005techniques,mutlu2006efficient,hashemi2016continuous,mutlu2005address,hashemi2015filtered,vector_runahead,mutlu2005reusing,iacobovici2004effective} and helper-thread execution~\cite{ssmt,precompute,software_preexecute,sohi_slice,helper_thread,bfetch,runahead_threads,helper_thread2,zhang2007accelerating,dubois1998assisted,collins2001speculative,solihin2002using}) pre-execute program code to generate future memory requests. These prefetchers can generate highly-accurate prefetches even when no \rbc{recognizable} pattern exists in \rbc{program} memory requests. However, \rbc{precomputation}-based prefetchers \rbc{usually} have high design complexity. Pythia \rbc{is not a} \rbc{precomputation}-based \rbc{proposal}. \rbc{It} finds patterns in past memory request \rbc{addressed} to generate prefetch \rbc{requests}.
\end{sloppypar}

Temporal prefetchers~\cite{markov,stems,somogyi_stems,wenisch2010making,domino,isb,misb,triage,wenisch2005temporal,chilimbi2002dynamic,chou2007low,ferdman2007last,hu2003tcp,bekerman1999correlated,cooksey2002stateless,karlsson2000prefetching} 
\rbc{memorize} long sequences of cacheline addresses \rbc{demanded by the processor}.
\rbc{When} a previously-seen address is encountered \rbc{again}, a temporal prefetcher \rbc{issues prefetch requests} \rbc{to} addresses that previously followed the currently-seen cacheline address. 
However, temporal prefetchers \rbc{usually} have \rbc{high storage requirements (often multi-megabytes of \rbc{metadata} storage, which necessitates storing metadata in memory~\cite{stems,somogyi_stems,isb}).}
Pythia requires only $25.5$KB of storage, \rbc{which can easily fit inside a core.}

Spatial prefetchers~\cite{stride,streamer,baer2,jouppi_prefetch,ampm,fdp,footprint,sms,spp,vldp,sandbox,bop,dol,dspatch,bingo,mlop,ppf,ipcp} predict a cacheline delta or \rbc{spatial} bit-pattern by learning program access patterns over \rbc{different} spatial \rbc{memory} \rbc{regions}. Spatial prefetchers provide high-accuracy prefetches, \rbc{usually} with \rbc{lower} storage overhead \rbc{than temporal prefetchers}. 
We already compare Pythia with other spatial prefetchers~\cite{mlop,bingo,spp,ppf,dspatch} and \rbc{show higher performance benefits}. 

\textbf{Machine Learning (ML) in Computer Architecture.}
\rbc{Prior works \rbc{apply} ML techniques in computer architecture in two \rbc{major} ways: (1) to design adaptive, data-driven algorithms, and (2) to explore \rbc{the large} microarchitectural design-space.}
\rbc{Researchers have proposed ML-based algorithms for various microarchitectural tasks like memory scheduling~\cite{rlmc,morse}, cache \rbc{management}~\cite{glider, imitation,rl_cache,teran2016perceptron,balasubramanian2021accelerating}, branch prediction~\cite{tarsa,branchnet,perceptron,garza2019bit,BranchNetV2_2020,Zouzias2021BranchPA,jimenez2016multiperspective,tarjanTaco2005,jimenez2003fast,jimenez2002neural}, \rbc{address translation~\cite{margaritovTranslationNIPS18}} and hardware prefetching~\cite{peled2018neural,peled_rl,hashemi2018learning,shineural,shi2019learning,shineural_asplos,zeng2017long}. 
\rbc{Pythia provides three key advantages over prior ML-based prefetchers. First, Pythia can learn to prefetch from multiple program features and system-level feedback information inherent to its design.
Second, Pythia can be customized online.
Third, Pythia incurs low hardware overhead.}
}
\rbc{Researchers have also explored ML techniques to explore the large microarchitectural design space, e.g., NoC design~\cite{fettes2018dynamic,zheng2019energy,lin2020deep,yin2020experiences,ebrahimi2012haraq,ditomaso2016dynamic,clark2018lead,ditomaso2017machine,van2018extending,yin2018toward}, chip placement optimization~\cite{mirhoseini2021}, hardware resource assignment for accelerators~\cite{kao2020confuciux}. 
These works are orthogonal to Pythia.}









\vspace{-0.5em}
\section{Conclusion} \label{conclusion}
We introduce Pythia, \rbc{the first} customizable prefetching framework that formulates prefetching as a reinforcement learning (RL) problem. Pythia autonomously learns to prefetch using multiple program features and system-level feedback \rbc{information} to predict memory accesses. 
Our extensive evaluations show that Pythia not only outperforms five \rbc{state-of-the-art} prefetchers but also provides robust performance \rbc{benefits} across a wide-range of \rbc{workloads} and system configurations.
\rbc{Pythia's benefits come with very modest area and power overheads.}
\rbc{We believe \rbc{and hope} that Pythia \rbc{would} encourage the next generation of data-driven autonomous prefetchers that automatically learn far-sighted prefetching policies by interacting with the system. Such prefetchers \rbc{can} not only improve performance and efficiency under a wide variety of workloads and system configurations, but also reduce the system architect's burden in designing sophisticated \rbc{prefetching mechanisms}.}
\begin{acks}
\rbc{\rbc{We thank} all SAFARI Research Group members, especially Skanda Koppula, for insightful feedback.
We acknowledge the generous gifts provided by our industrial partners: Google, Huawei, Intel, Microsoft, and VMware.}
\rbc{This work is supported in part by the Semiconductor Research Corporation and the ETH Future Computing Laboratory.}
The first author thanks his departed father, whom he lost in COVID-19 pandemic.
\end{acks}

\bibliographystyle{ACM-Reference-Format}
\bibliography{refs}

\appendix
\section{Artifact Appendix} \label{sec:artifact}

\subsection{Abstract}

We implement Pythia using ChampSim simulator~\cite{champsim}. In this artifact, we provide the source code of Pythia and necessary instructions to reproduce its key performance results. We identify four key results to demonstrate Pythia's novelty: 
\begin{itemize}
    \item Workload category-wise performance speedup of all competing prefetchers (Fig.~\ref{fig:perf_1T}(a)).
    \item Workload category-wise coverage and overpredictions of all competing prefetchers (Fig.~\ref{fig:cov_acc}).
    \item Geomean performance comparison with varying DRAM bandwidth from $150$-MTPS to $9600$-MTPS (Fig.~\ref{fig:scalability_master}(b)).
    \item Workload category-wise performance speedup of all competing prefetchers (Fig.~\ref{fig:perf_4T}(a)).
\end{itemize}


The artifact can be executed in any machine with a general-purpose CPU and $52$ GB disk space. However, we strongly recommend running the artifact on a compute cluster with \texttt{slurm}~\cite{slurm} support for bulk experimentation.

\subsection{Artifact Check-list (Meta-information)}


\small
\begin{itemize}
  \item {\bf Compilation: } G++ v6.3.0 or above.
  \item {\bf Data set: } Download traces using the supplied script.
  \item {\bf Run-time environment: } Perl v5.24.1
  \item {\bf Metrics: } IPC, prefetcher's coverage, and accuracy.
  \item {\bf Experiments: } Generate experiments using supplied scripts.
  \item {\bf How much disk space required (approximately)?: } $52$GB
  \item {\bf How much time is needed to prepare workflow (approximately)?: } $\sim 2$ hours. Mostly depends on the time to download traces.
  \item {\bf How much time is needed to complete experiments (approximately)?: } 3-4 hours using a compute cluster with $480$ cores.
  \item {\bf Publicly available?: } Yes.
  \item {\bf Code licenses (if publicly available)?: } MIT
  \item {\bf Archived (provide DOI)?: } \url{https://doi.org/10.5281/zenodo.5520125}
\end{itemize}

\subsection{Description}

\subsubsection{How to Access}

The source code can be downloaded either from GitHub (\url{https://github.com/CMU-SAFARI/Pythia}) or from Zenodo (\url{https://doi.org/10.5281/zenodo.5520125}).

\subsubsection{Hardware Dependencies}

Pythia can be run on any system with a general-purpose CPU and at least $52$ GB of free disk space.

\subsubsection{Software Dependencies}
\begin{sloppypar}
Pythia requires \texttt{GCC v6.3.0} and \texttt{Perl v5.24.1}. Optionally, Pythia requires \texttt{megatools v1.9.98} to download few traces, and Microsoft Excel (tested on v16.51) to reproduce results as presented in the paper.
\end{sloppypar}

\subsubsection{Data Sets}

The ChampSim traces required to evaluate Pythia can be downloaded using the supplied script. Our implementation of Pythia is fully compatible with prior ChampSim traces that are used in previous cache replacement (CRC-2~\cite{crc2}), data prefetching (DPC-3~\cite{dpc3}) and value-prediction (CVP-2~\cite{cvp2}) championships. We are also releasing a new set of ChampSim traces extracted from Ligra~\cite{ligra} and PARSEC-2.1~\cite{parsec} suites.


\subsection{Installation}

\begin{enumerate}
    \item Clone Pythia from GitHub repository:
        \vspace{0.4em}
        \shellcmd{git clone https://github.com/CMU-SAFARI/Pythia.git}
        \vspace{0.4em}
    
    \item Clone Bloomfilter library inside Pythia home and build:
        \vspace{0.4em}
        \shellcmd{cd Pythia/}
        \shellcmd{git clone https://github.com/mavam/libbf.git libbf/}
        \shellcmd{cd libbf/}
        \shellcmd{mkdir build \&\& cd build/ \&\& cmake ../}
        \shellcmd{make clean \&\& make}
        \vspace{0.4em}
        
    \item Build Pythia for single-core and four-core configurations:
        \vspace{0.4em}
        \shellcmd{cd \$PYTHIA\_HOME}
        \shellcmd{./build\_champsim.sh multi multi no 1}
        \shellcmd{./build\_champsim.sh multi multi no 4}
        \vspace{0.4em}
    
    \item Please make sure to set environment variables as:
        \vspace{0.4em}
        \shellcmd{source setvars.sh}
\end{enumerate}

\subsection{Experiment Workflow}
This section describes steps to generate, and execute necessary experiments. We recommend the reader to follow the README file to know more about each script used in this section.

\subsubsection{Preparing Traces}
\begin{enumerate}
    \item Download necessary traces as follows:
        \vspace{0.4em}
        \shellcmd{mkdir \$PYTHIA\_HOME/traces}
        \shellcmd{cd \$PYTHIA\_HOME/scripts}
        \shellcmd{perl download\_traces.pl --csv artifact\_traces.csv \\ --dir \$PYTHIA\_HOME/traces/}
        \vspace{0.4em}
    \item If the traces are downloaded in other path, please update the full path in \texttt{MICRO21\_1C.tlist} and \texttt{MICRO21\_4C.tlist} inside \\ \texttt{\$PYTHIA\_HOME/experiments} directory appropriately.
\end{enumerate}

\subsubsection{Launching Experiments}\label{sec:artifact_launching_experiments}
The following instructions will launch all experiments required to reproduce key results in a local machine. We \textbf{strongly} recommend using a compute cluster with \texttt{slurm} support to efficiently launch experiments in bulk. To launch experiments using \texttt{slurm}, please provide \texttt{--local 0} (tested using \texttt{slurm v16.05.9}). 
\begin{enumerate}
    \item Launch single-core experiments as follows:
        \vspace{0.4em}
        \shellcmd{cd \$PYTHIA\_HOME/experiments}
        \shellcmd{perl \$PYTHIA\_HOME/scripts/create\_jobfile.pl --exe \\ \$PYTHIA\_HOME/bin/perceptron-multi-multi-no-ship-1core \\--tlist MICRO21\_1C.tlist --exp MICRO21\_1C.exp --local 1 > jobfile.sh}
        \shellcmd{cd experiments\_1C}
        \shellcmd{source ../jobfile.sh}
        \vspace{0.4em}
    \item Launch four-core experiments as follows:
        \vspace{0.4em}
        \shellcmd{cd \$PYTHIA\_HOME/experiments}
        \shellcmd{perl \$PYTHIA\_HOME/scripts/create\_jobfile.pl --exe \\ \$PYTHIA\_HOME/bin/perceptron-multi-multi-no-ship-4core \\--tlist MICRO21\_4C.tlist --exp MICRO21\_4C.exp --local 1 > jobfile.sh}
        \shellcmd{cd experiments\_4C}
        \shellcmd{source ../jobfile.sh}
        \vspace{0.4em}
    \item Please make sure the paths used in \texttt{tlist} and \texttt{exp} files are appropriately changed before creating the experiment files.
\end{enumerate}

\subsubsection{Rolling-up Statistics}
We will use \texttt{rollup.pl} script to rollup statistics from outputs of all experiments. To automate the process, we will use the following instructions. This will create three comma-separated-value (CSV) files in experiments directory which will be used for evaluation in \cref{sec:artifact_evaluation}.
\vspace{0.4em}
\shellcmd{cd \$PYTHIA\_HOME/experiments}
\shellcmd{bash automate\_rollup.sh}


\subsection{Evaluation}\label{sec:artifact_evaluation}
For single-core baseline configuration experiments, we will evaluate three metrics: performance, coverage, and overprediction of each prefetcher. For single-core experiments varying DRAM-bandwidth and four-core experiments, we will evaluate only performance. The performance, coverage and overprediction of a prefetcher X is measured by following equations:
\[ Perf_X = \frac{IPC_X}{IPC_{nopref}} \]
\[ Coverage_X = \frac{LLC\_load\_miss_{nopref}-LLC\_load\_miss_X}{LLC\_load\_miss_{nopref}} \]
\[ Overprediction_X = \frac{LLC\_read\_miss_{X}-LLC\_read\_miss_{nopref}}{LLC\_read\_miss_{nopref}} \]

To easily calculate the metrics, we are providing a Microsoft Excel template to post-process the rolled-up CSV files. The template has four sheets, three of which bear the same name of the rolled up CSV files. Each sheet is already populated with our collected results, necessary formulas, pivot tables, and charts to reproduce the results presented in the paper.
Please follow the instructions to reproduce the results from your own CSV statistics files:
\begin{enumerate}
    \item Copy and paste each CSV file in the corresponding sheet's top left corner (i.e., cell A1).
    \item Immediately after pasting, convert the comma-separated rows into columns by going to Data -> Text-to-Coloumns -> Selecting comma as a delimiter. This will replace the already existing data in the sheet with the newly collected data.
    \item \emph{Refresh each pivot table in each sheet} by clicking on them and then clicking Pivot-Table-Analyse -> Refresh.
\end{enumerate}
The reader can also use any other data processor (e.g., Python pandas) to reproduce the same result. 

\subsection{Expected Results}
\begin{itemize}
    \item In single-core experiments, Pythia should achieve $22\%$ performance improvement over the no prefetching baseline, with $71\%$ prefetch coverage and $27$\% overpredictions. The next best prefetcher MLOP should achieve $19$\% performance improvement, with $64$\% coverage and $110$\% overpredictions.
    \item In single-core experiments with DRAM bandwidth scaling, Pythia should achieve the following:
    \begin{itemize}
        \item In $150$-MTPS configuration, Pythia should achieve $0.89$\% performance improvement over no prefetching baseline, whereas MLOP should \emph{underperform} baseline by $16$\%.
        \item In $9600$-MTPS configuration, Pythia should achieve $22.7$\% performance improvement over no prefetching baseline, whereas MLOP should achieve $19.5$\%.
    \end{itemize}
    \item In four-core experiments, Pythia should achieve $30$\% performance improvement over no prefetching baseline, whereas MLOP should achieve $24$\%. 
\end{itemize}

\subsection{Experiment Customization}
\begin{itemize}
    \item The configuration of every prefetcher can be customized by changing the \texttt{ini} files inside the \texttt{config} directory.
    \item The \texttt{exp} files can be customized to run new experiments with different prefetcher combinations. More experiment files can be found inside \texttt{experiments/extra} directory. One can use the same instructions mentioned in \cref{sec:artifact_launching_experiments} to launch experiments.
\end{itemize}

\subsection{Methodology}

Submission, reviewing and badging methodology:

\begin{itemize}
  \item \url{https://www.acm.org/publications/policies/artifact-review-badging}
  \item \url{http://cTuning.org/ae/submission-20201122.html}
  \item \url{http://cTuning.org/ae/reviewing-20201122.html}
\end{itemize}
\section{Extended Results}

\subsection{Detailed Performance Analysis}
\subsubsection{\textbf{Single-core}}
\begin{figure}[!h]
\centering
\includegraphics[width=3.3in]{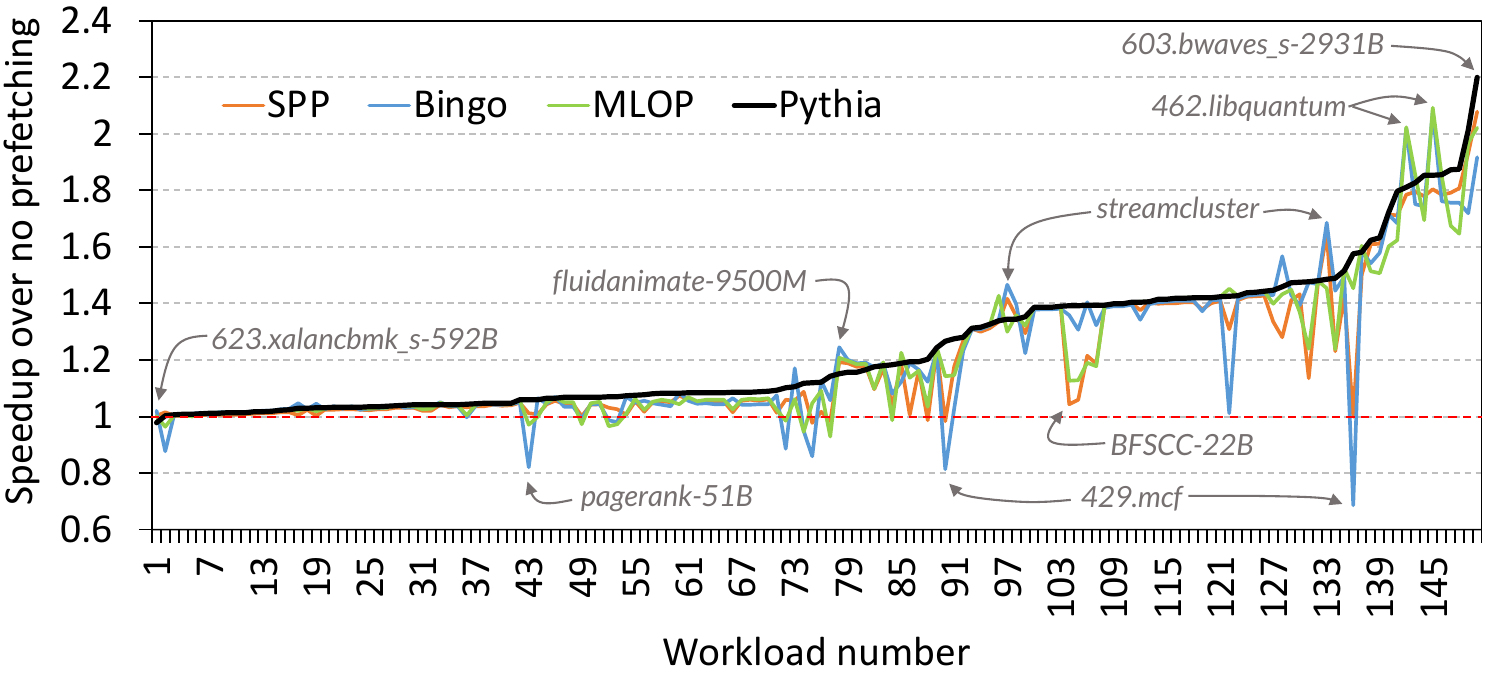}
\caption{Performance line graph of 150 single-core traces.}
\label{fig:perf_1T_scurve}
\end{figure}

\begin{sloppypar}
Fig.~\ref{fig:perf_1T_scurve} shows the performance line graph of all prefetchers \rbcc{for the $150$} single-core workload traces. The workload traces are sorted in ascending order of performance improvement \rbcc{of} Pythia over the baseline without prefetching. We make three key observations. First, Pythia outperforms the no-prefetching baseline in every single-core trace, except \texttt{623.xalancbmk-592B} (where it underperforms the baseline by $2.1$\%). \texttt{603.bwaves-2931B} enjoys the highest performance improvement of $2.2\times$ over the baseline. \rbcc{Performance of the} top $80$\% of traces improve by at least $4.2$\% \rbcc{over the baseline}. Second, Pythia underperforms Bingo in workloads like \texttt{libquantum} 
due to \rbcc{the heavy streaming} nature of memory accesses. As \texttt{libquantum} streams through all physical pages, Bingo simply prefetches all cachelines of a page \rbcc{at once} just by seeing the first access \rbcc{to} the page. As a result Bingo achieves higher timeliness and higher performance than Pythia. Third, Pythia significantly outperforms every competing \rbcc{prefetcher} in workloads with irregular access patterns (e.g., \texttt{mcf}, \texttt{pagerank}).
We conclude that Pythia provides consistent performance \rbcc{gains} over the no-prefetching baseline and multiple prior state-of-the-art prefetchers over a wide range of workloads. We share a table \rbcc{depicting} the single-core performance of every competing prefetcher considered in this paper over the no-prefetching baseline in our GitHub repository: \url{https://github.com/CMU-SAFARI/Pythia}.
\end{sloppypar}

\subsubsection{\textbf{Four-core}}
\begin{figure}[!h]
\centering
\includegraphics[width=3.3in]{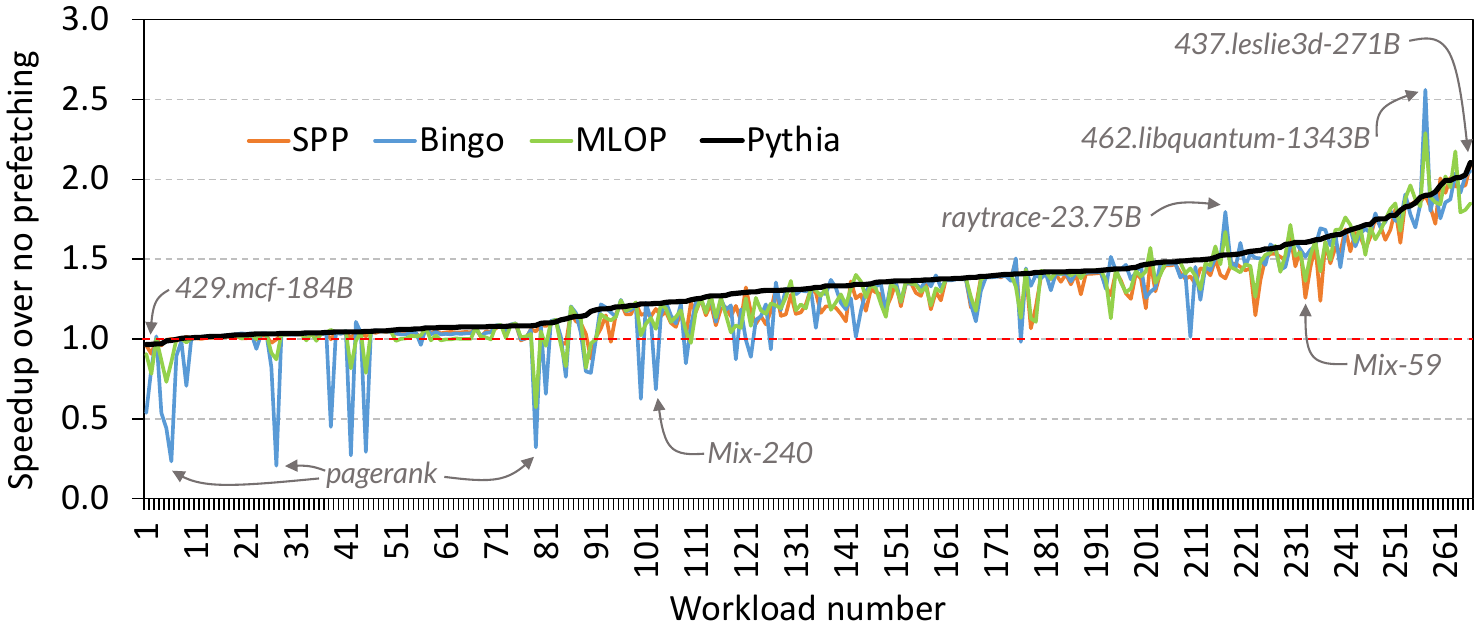}
\caption{Performance line graph of 272 four-core trace mixes.}
\label{fig:perf_4T_scurve}
\end{figure}

Fig.~\ref{fig:perf_4T_scurve} shows the performance line graph of all prefetchers \rbcc{for 272} four-core workload trace mixes (including both homogeneous and heterogeneous mixes). The workload mixes are sorted in ascending order of performance improvement \rbcc{of} Pythia over the baseline without prefetching. We make two key observations. First, Pythia outperforms the baseline without prefetching in \rbcc{all but one} four-core trace mix. Pythia provides the highest performance gain in \texttt{437.leslie3d-271B} ($2.1\times$) and lowest performance gain in \texttt{429.mcf-184B} (-$3.5$\%) over the no-prefetching baseline. Second, Pythia also outperforms (or matches performance) all competing prefetchers in \rbcc{majority of} trace mixes. Pythia underperforms Bingo in \rbcc{the} \texttt{462.libquantum} homogeneous trace mix due to \rbcc{the very} regular streaming access pattern. On the other hand, Pythia significantly outperforms \rbcc{Bingo} in \texttt{Ligra} workloads (e.g., \texttt{pagerank}) due to its adaptive prefetching strategy to trade-off coverage for accuracy in high memory bandwidth usage. We conclude that Pythia provides a consistent performance gain over multiple prior state-of-the-art prefetchers over a wide range of workloads even in bandwidth-constrained \rbcc{multi-core} systems.

\subsection{Performance with Different Features}
\begin{figure}[!h]
\centering
\includegraphics[width=\columnwidth]{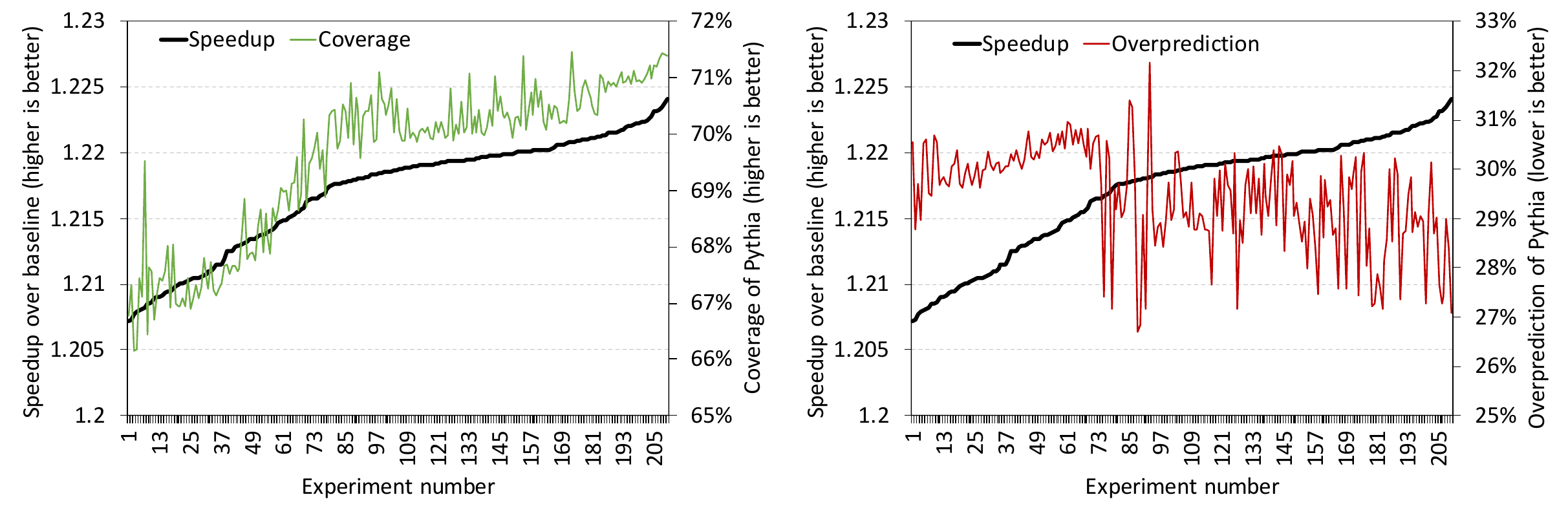}
\caption{Performance, coverage, and overprediction of Pythia with different feature combinations. \rbcc{The x-axis shows experiments with different feature combinations.}}
\label{fig:feature_sel}
\end{figure}

Fig.~\ref{fig:feature_sel} shows the performance, coverage, and overprediction of Pythia averaged across all single-core traces with different feature combinations during automated feature selection (\cref{sec:tuning_feature_selection}). For brevity, we show results for all experiments with any-one and any-two combinations of $20$ features taken from the full list of $32$ features. Both graphs are sorted in ascending order of performance improvement \rbcc{of Pythia} over the baseline without prefetching. We make three key observations. First, Pythia's performance gain over the no-prefetching baseline improves from $20.7$\% to $22.4$\% \rbcc{by varying} the feature combination. We select the feature combination that provides the highest performance gain as the basic Pythia configuration (Table~\ref{tab:pythia_config}). Second, Pythia's coverage and overprediction also change significantly with \rbcc{varying} feature \rbcc{combination}. Pythia's coverage improves from $66.2$\% to $71.5$\%, whereas overprediction improves from $32.2$\% to $26.7$\% by changing feature \rbcc{combination}. Third, Pythia's performance gain positively correlates \rbcc{with} Pythia's coverage in single-core configuration. \rbcc{We conclude that automatic design-space exploration can significantly optimize Pythia's performance, coverage, and overpredictions.}

\subsection{Performance Sensitivity to Hyperparameters}
\begin{figure}[!h]
\centering
\includegraphics[width=3.3in]{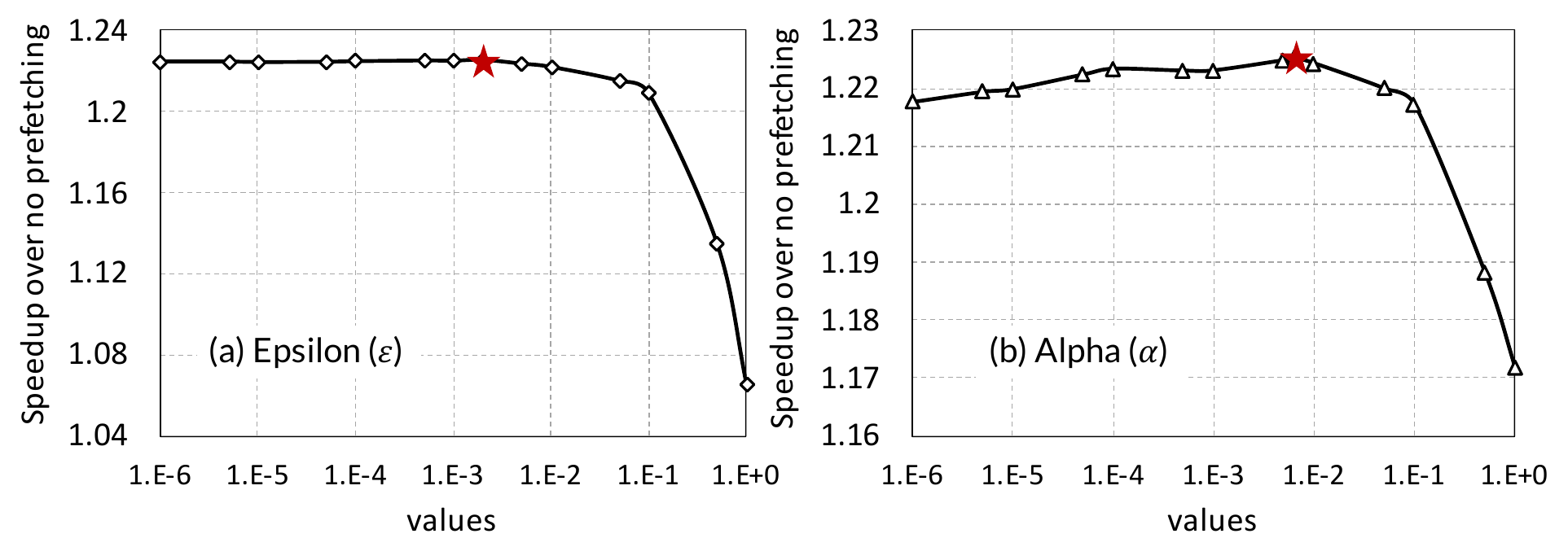}
\caption{Performance sensitivity of Pythia towards (a) the exploration rate ($\epsilon$), and (b) the learning rate ($\alpha$) hyperparameter values. The values in basic Pythia configuration are marked in red.}
\label{fig:hyp_sensitivity}
\end{figure}

Fig.~\ref{fig:hyp_sensitivity}(a) shows Pythia's performance sensitivity \rbcc{to the} exploration rate ($\epsilon$) averaged across all single-core traces. The key takeaway from \rbcc{Figure} \ref{fig:hyp_sensitivity}(a) is that Pythia's performance improvement drops sharply if the underlying RL-agent heavily \emph{explores} the state-action space \rbcc{as opposed to exploiting the learned policy}. Changing the $\epsilon$-value from $0.002$ to $1.0$ \rbcc{reduces Pythia's} performance improvement by $16.0$\%.
Fig.~\ref{fig:hyp_sensitivity}(b) shows Pythia's performance sensitivity \rbcc{to} learning rate \rbcc{parameter} ($\alpha$), averaged across all single-core traces. The key takeaway from Fig.~\ref{fig:hyp_sensitivity}(b) is \rbcc{that Pythia's performance improvement reduces for both increasing or decreasing the learning rate parameter. Increasing the learning rate reduces the hysteresis in Q-values (i.e., Q-values change significantly with the immediate reward received by Pythia), which reduces Pythia's performance improvement. Similarly, decreasing the learning rate also reduces Pythia's performance as it increases the hysteresis in Q-values.}
Pythia achieves optimal performance improvement for $\alpha=0.0065$. 

\subsection{Comparison to the Context Prefetcher}
\begin{figure}[!h]
\centering
\includegraphics[width=\columnwidth]{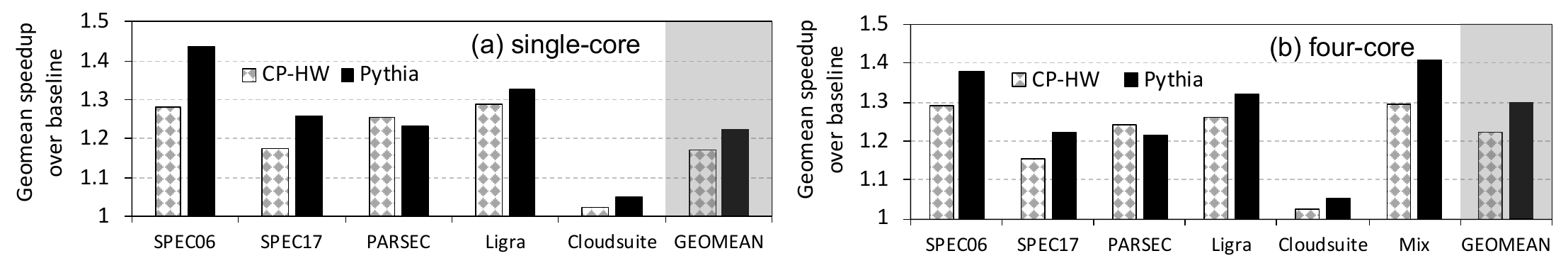}
\caption{Performance of Pythia vs. the context prefetcher~\cite{peled_rl} using hardware contexts.}
\label{fig:perf_context_pref}
\end{figure}

As we discuss in Section~\ref{sec:compare_with_context}, unlike Pythia, the context prefetcher (CP~\cite{peled_rl}) relies on both hardware and software contexts. A tailor-made compiler needs to encode the software contexts using \rbcc{special} NOP instructions, which are decoded by the core front-end to pass the context to the CP.
For a fair comparison, we implement the context prefetcher using hardware contexts (CP-HW) and show the performance comparison \rbcc{of Pythia and CP-HW} in Figure~\ref{fig:perf_context_pref}. \rbcc{The key takeaway is that} Pythia outperforms the CP-HW prefetcher by $5.3$\% and $7.6$\% in single-core and four-core configurations, respectively.  \rbcc{Pythia's performance improvement over CP-HW mainly comes from two key aspects: (1) Pythia's ability to take memory bandwidth usage into consideration while taking prefetch actions, and (2) the far-sighted predictions made by Pythia as opposed to myopic predictions by CP-HW.}

\subsection{Comparison to \rbcc{the IBM POWER7} Adaptive Prefetcher}
\begin{figure}[!h]

\centering
\includegraphics[width=\columnwidth]{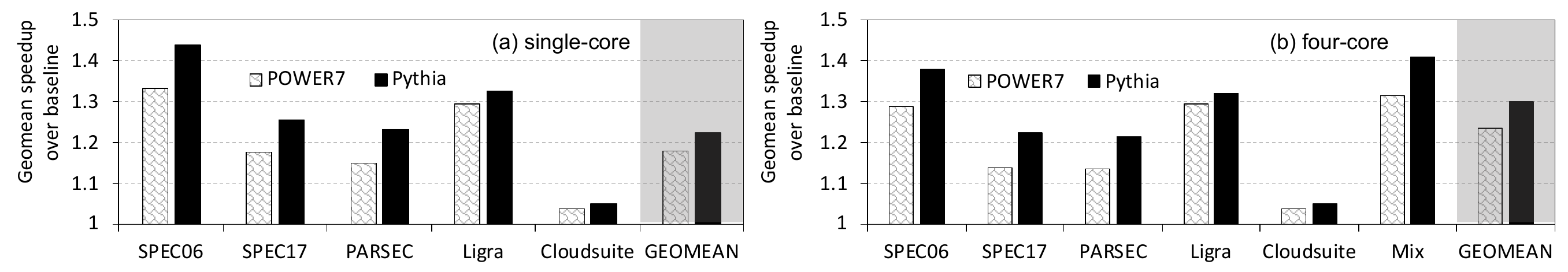}
\caption{Performance comparison against IBM POWER7 prefetcher~\cite{ibm_power7}.}
\label{fig:perf_power7}
\end{figure}

\rbcc{Fig.~\ref{fig:perf_power7} compares Pythia against the IBM POWER7 adaptive prefetcher~\mbox{\cite{ibm_power7}}. The POWER7 prefetcher dynamically tunes its prefetch aggressiveness (e.g., selecting prefetch depth, enabling stride-based prefetching) by monitoring program performance.}
We make two observations from Fig.~\ref{fig:perf_power7}. First, Pythia outperforms \rbcc{the} POWER7 prefetcher by $4.5$\% in single-core system. This is mostly due to Pythia's ability \rbcc{to} \rbcc{capture} different types of address patterns than just streaming/stride patterns. Second, Pythia outperforms POWER7 prefetcher by $6.5$\% in four-core and $6.1$\% in eight-core systems (not \rbcc{plotted}), respectively. The increase in performance improvement from single to four (or eight) core configuration suggests \rbcc{that} Pythia \rbcc{is more adaptive} than \rbcc{the} POWER7 prefetcher.

\subsection{Performance Sensitivity to Number of Warmup Instructions}
\begin{figure}[!h]
\vspace{-0.5em}
\centering
\includegraphics[width=\columnwidth]{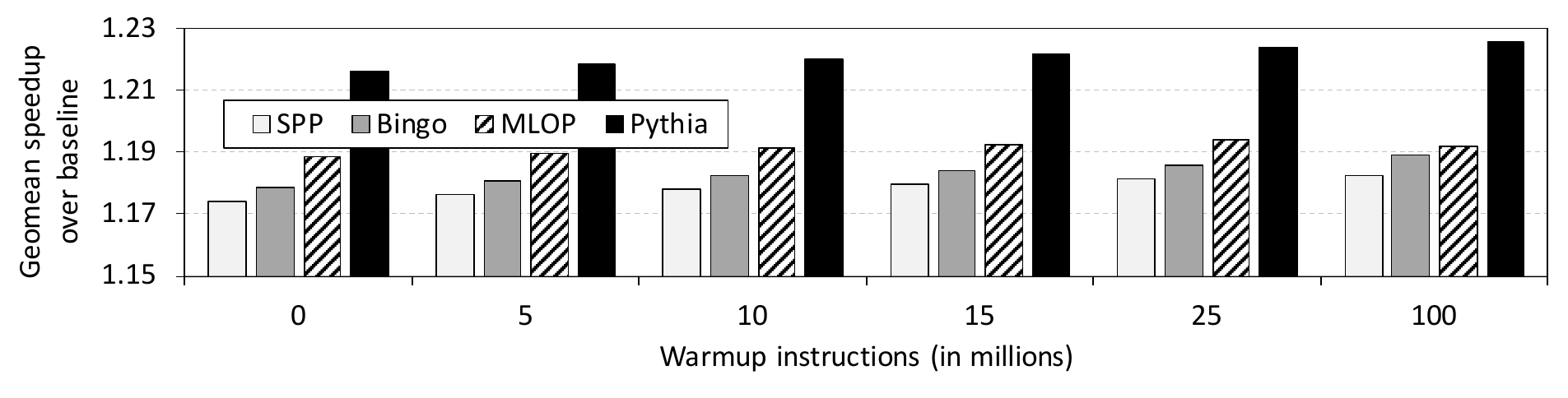}
\vspace{-2em}
\caption{Performance sensitivity of all prefetchers to number of warmup instructions.}
\label{fig:warmup_sen}
\vspace{-1em}
\end{figure}

\rbcd{Fig.~\ref{fig:warmup_sen} shows performance sensitivity of all prefetchers \rbcc{to the} number of warmup instructions averaged across all single-core traces. Our baseline simulation configuration uses $100$ million warmup instructions. The key takeaway from Fig.~\ref{fig:warmup_sen} is that Pythia consistently outperforms prior prefetchers in a wide range of simulation configurations using different number of warmup instructions. In the baseline simulation configuration using $100$M warmup instructions, Pythia outperforms MLOP, Bingo, and SPP by $3.4$\%, $3.8$\%, and $4.4$\% respectively. In a simulation configuration with no warmup instruction, Pythia continues to outperform MLOP, Bingo, and SPP by $2.8$\%, $3.7$\%, and $4.2$\% respectively. We conclude that, Pythia can quickly learn to prefetch from a program's memory access pattern and provides higher performance than other heuristics-based prefetching techniques over a wide range of simulation configurations using different number of warmup instructions.}

\end{document}